\documentclass[11pt]{article}

\usepackage[usenames,dvipsnames,table]{xcolor}

\pdfoutput=1 
\usepackage{jheppub} 

\usepackage{graphicx}
\usepackage{amsmath, amssymb}

\newcommand{\be}{\begin{eqnarray}}
\newcommand{\ee}{\end{eqnarray}}
\newcommand{\bea}{\begin{eqnarray}}
\newcommand{\eea}{\end{eqnarray}}

\newcommand{\bn}{\begin{enumerate}}
\newcommand{\en}{\end{enumerate}}

\def\half{\frac{1}{2}}

\title{From $6d$ flows to $4d$ flows}

\preprint{}

\author[a]{Shlomo S. Razamat,}
\author[a]{Evyatar Sabag,}
\author[b]{and Gabi Zafrir}

\affiliation[a]{Department of Physics, Technion, Haifa, 32000, Israel}
\affiliation[b]{Kavli IPMU (WPI), UTIAS, the University of Tokyo, Kashiwa, Chiba 277-8583, Japan}

\emailAdd{razamat@physics.technion.ac.il}
\emailAdd{sevyatar@campus.technion.ac.il}
\emailAdd{gabi.zafrir@ipmu.jp}

\abstract{SCFTs in six dimensions are interrelated by networks of RG flows. Compactifying such models on a Riemann surface with flux for the $6d$ global symmetry, one can obtain a wide variety of theories in four dimensions. These four dimensional models are also related by a network of RG flows. In this paper we study some examples of four dimensional flows relating theories that can be obtained from six dimensions starting with different SCFTs connected by $6d$ RG flows. We compile a dictionary between different orders of such flows, $6d\to 6d\to 4d$ and $6d\to 4d\to 4d$, in the particular case when the six dimensional models are the ones residing on M5 branes probing different $A$-type singularities. The flows we study are triggered by vacuum expectation values (vevs) to certain operators charged under the six dimensional symmetry.
We find that for generic choices of parameters the different orders of flows, $6d\to 6d\to 4d$ and $6d\to 4d\to 4d$, involve compactifications on  different Riemann surfaces with the difference being in the number of punctures the surface has. 

}

\begin{document} 

\maketitle
\flushbottom

\section{Introduction}

A huge variety of conformal field theories (CFTs) can be engineered as a description of a fixed point of an RG flow.  Defining an RG flow involves choosing a UV CFT and a deformation breaking the conformal symmetry.  Different starting points and deformations can lead to the same IR fixed point CFT. An interesting question in trying to understand the space of possible conformal theories, is to understand such equivalence classes of flows. 

\begin{figure}[t]
	\centering
  	\includegraphics[scale=0.64]{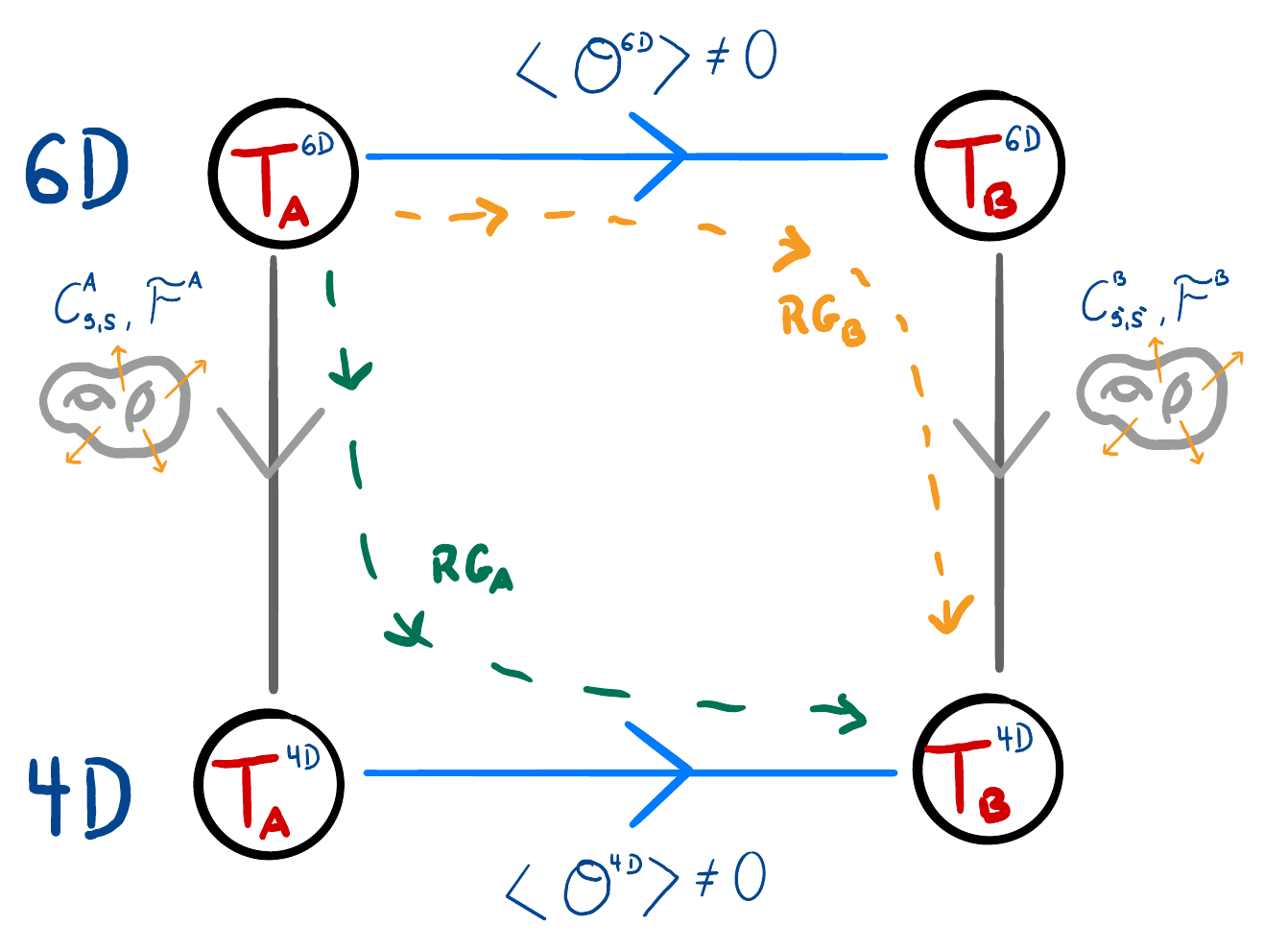}
    \caption{A diagram representing different RG flows we are considering. Flow $RG_A$ describes a compactification of a $6d$ model to an effective $4d$ theory followed by turning on a vacuum expectation value for an operator in $4d$. Flow $RG_B$ is defined by first turning on a vacuum expectation value for a $6d$ operator and then compactifying the IR model to $4d$. The question we will discuss in what follows is given one of the two flows, $RG_A$ or $RG_B$, how to identify the other one. }
    \label{F:generalflows}
\end{figure}

In recent years, a large amount of four dimensional supersymmetric conformal field theories have been engineered as a low energy description of six dimensional theories, with two dimensions being a compact Riemann surface \cite{Gaiotto:2009we,Bah:2012dg,Gaiotto:2015usa,Kim:2017toz,Kim:2018lfo,Razamat:2016dpl,Kim:2018bpg,Razamat:2018gro,Apruzzi:2018oge,Heckman:2016xdl,Chen:2019njf}.  One can think of such a construction as yet another construction of an RG flow leading to a CFT. In this case the UV CFT is a six dimensional theory and the relevant deformation is the geometry itself. Below the energy scale set by the size of the compact part of the geometry we obtain an effective four dimensional model which might flow to an interesting CFT.
To engineer a theory in four dimensions in this manner we have several choices. One of them is the choice of theory in six dimensions (see \cite{Heckman:2018jxk} for a review of $6d$ SCFTs) and another is a collection of choices related to the compactification. Once a theory in four dimensions is engineered, further deformations can be preformed by either turning on relevant interactions or vacuum expectation values (vevs). The same model again might be reached starting from different choices in six dimensions and different flows (see Fig. \ref{F:generalflows} for illustration). A very interesting question is then to understand such equivalence classes of flows in this restricted but very rich set of theories.

We will consider in this paper two types of sequences of flows. We will denote the first as $6d\to 4d\to 4d$, and it starts from a $6d$ SCFT, first compactified to a four dimensional model followed by another purely $4d$ flow triggered by a vev to a four dimensional operator. This flow is labeled by a choice of compactification data and the $4d$ operator, $RG_A=((\Sigma_{g,s}, {\cal F})_A, {\cal O}_{4d})$.
Here $\Sigma_{g,s}$ is the choice of the compactification surface, which we take to be a genus $g$ Riemann surface with $s$ punctures, and ${\cal F}$ is the choice of flux for the global symmetry in six dimensions. We denote the second flow by $6d\to 6d\to4d$, and it also starts from a $6d$ SCFT, but first flows to a different $6d$ SCFT by a vev to a $6d$ operator followed by a geometric flow through compactification to $4d$. Such a flow is labeled by a choice of $6d$ operator and the compactification geometry, $RG_B=((\Sigma_{g',s'}, {\cal F}')_B, {\cal O}_{6d})$. A natural question we will consider is which choices of $RG_A$ and $RG_B$ lead to the same fixed point in four dimensions. 

We will develop such a dictionary in the particular case where the six dimensional models are residing on M5 branes probing a $\mathbb{Z}_k$ singularity denoted by $\mathcal{T}(SU(k),N)$, and the $6d$ flows are flows between these models.
Such models and their compactifications were extensively studied recently \cite{DelZotto:2014hpa,Ohmori:2014kda,Gaiotto:2015usa} and we have a rich enough set of tools to study the flows between them. In particular, we will discover that by choosing naively similar vevs in six and four dimensions, the compactification geometry leading to equivalent models in the two flows does not have to be the same, and in fact it often differs by the number of punctures. The difference in the number of punctures, in particular, depends on details of the flux. We will employ several computational tools to arrive at our conclusions. 

One such tool is a simple limit of the supersymmetric index which we have at our disposal for the particular theories of interest \cite{Razamat:2018zus}. This limit generalizes the Coulomb branch limit of the ${\cal N}=2$ index \cite{Gadde:2011uv} to ${\cal N}=1$ models obtained from compactifications of SCFTs residing on M5 branes probing A-type orbifold singularities. Such a limit of the index captures very limited but non trivial information about the CFTs on one hand, and can be computed for almost arbitrary compactification choices on the other hand. For the particular cases of compactifications of $\mathcal{T}(SU(2),2)$ and compactification on tori of $\mathcal{T}(SU(k),N)$, we will be able to make use of the full index.  Another set of tools is given by 't Hooft anomaly matching both for $4d$ and $6d$ flows.

The outline of the paper is as follows. In Section \ref{S:6dModels} we will discuss the generalities of the six dimensional models and their possible flows. In Section \ref{S:TaleOf2Flows} we will discuss the two sequences of flows $6d \rightarrow 6d \rightarrow 4d$ and $6d \rightarrow 4d \rightarrow 4d$ and explain how to settle their apparent discrepancy. In Section \ref{S:6d flow} we will derive results regarding the $6d$ flow from $\mathcal{T}\left(SU(k),N\right)$ to $\mathcal{T}\left(SU(k-1),N\right)$. In Section \ref{S:Coulomb limit} we consider the Coulomb limit of the index and in Section \ref{S:IndexComp} the full index in the cases it is known. In Section \ref{S:APFlow} we will study the flows at the level of the $4d$ anomaly polynomial.  Several appendices complement the bulk of the manuscript with additional technical details.

\section{$6d$ SCFTs operators and flows}
\label{S:6dModels}

\begin{figure}[t]
	\centering
  	\includegraphics[scale=0.3]{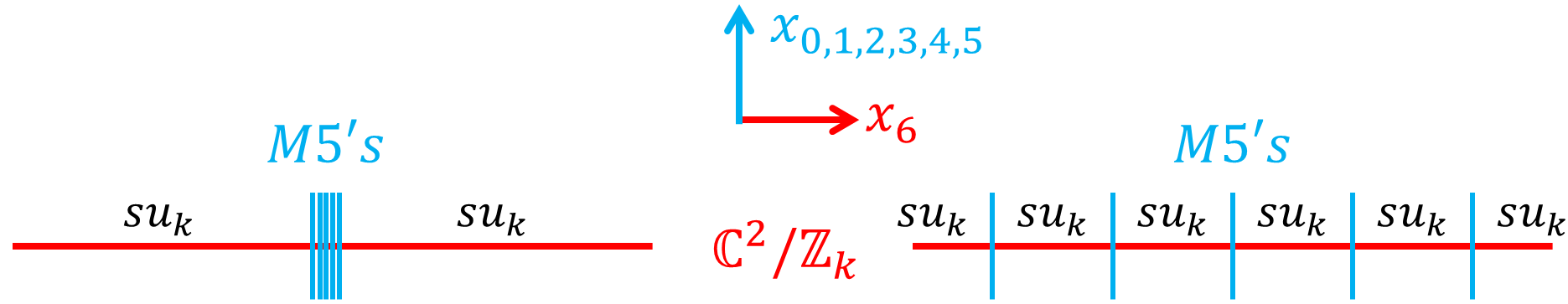}
    \caption{An M-theory brane construction of M5 branes on top of a $\mathbb{C}^2/\mathbb{Z}_k$ singularity occupying directions $7,8,9,10$. The M5-branes spread in directions $0,1,2,3,4,5$. On the left: The SCFT point is found when all the M5-branes lay on top of one another. On the right: The M5-branes are separated along direction $6$ giving the tensor branch description that can be interpreted as a quiver description of $SU(k)$ nodes connected with bifundamental hypermultiplets.}
    \label{F:M5Sing}
\end{figure}

The $6d$ models we will use in this manuscript are ones described by the IR behavior of a stack of M5-branes probing a $\mathbb{C}^2/\mathbb{Z}_k$ singularity. The models are denoted by $\mathcal{T}\left(SU(k),N\right)$, where $k$ comes from the singularity, and $N$ is the number of probing M5-branes. These models and many other $6d$ SCFTs can be found in \cite{DelZotto:2014hpa,Heckman:2015bfa}. The SCFT points of these models are found when all the M5-branes lay on top of one another (Fig. \ref{F:M5Sing} left). At this point effective strings living on the M5-branes become tensionless and the theory has no scale. These effective strings are M2-branes sharing one spatial direction and time with the M5-branes, and stretch in another orthogonal direction between different M5-branes. The SCFT point is strongly coupled and very little is known about it; therefore, we use the tensor branch description of the theory. In this description all the M5-branes are separated along the line of the singularity and we can find a quiver description of the theory (Fig. \ref{F:M5Sing} right). In the field theory we give a vacuum expectation value to the scalars of the tensor multiplets to find the tensor branch description.
The tensor branch allows us to understand the quiver description of the theory. From the quiver we can read off the global symmetry of the theory, and the matter, gauge and tensor multiplets that build it. In the SCFT point the symmetry may be enhanced or restricted. Most importantly, 't Hooft anomalies can be calculated on the tensor branch using the techniques of \cite{Ohmori:2014kda}, and are expected to remain unchanged in the SCFT point. This will prove very useful when dealing with the theories described above.

The flavor symmetries of these models on the tensor branch are $SU(k)_\beta \times SU(k)\gamma \times U(1)$. The $SU(k)$ symmetries come from the half infinite segments with symmetry $SU(k)$. The $U(1)$ is the symmetry acting identically on all the bifundamental hypermultiplets. Naively, there are $N$ $U(1)$ groups, each acting on a different hypermuliplet, however, ABJ type gauge-flavor anomalies break all of them save for the  diagonal combination.  In some specific cases this symmetry is enhanced. When $N=2$ the symmetry is enhanced to $SU(2k)\supset SU(k)_\beta \times SU(k)\gamma \times U(1)_t$, which is also visible in the tensor branch quiver. When $k=2$ the $U(1)_t$ symmetry is enhanced to $SU(2)_t$ symmetry, and when both $N=k=2$ the flavor symmetry is enhanced to $SO(7)$. In the tensor branch quiver description this enhancement is due to the $SU(2)\times SU(2)$ bifundamental being a real representation, and so rotated with an $SU(2)$ group rather than a $U(1)$. Nevertheless, the symmetry of the SCFT in these cases is smaller than that expected from the quiver. One reason for this is that the anomalies that break the $N$ $U(1)$ groups to the diagonal do not exist for the $SU(2)$ case so naively each bifundamental hyper can be rotated separately. The reduction of the global symmetry at the SCFT point was first suggested in \cite{Ohmori:2015pia} for the $N=k=2$ case, and based on it one can also argue the symmetry reduction in the $N>2$ cases \cite{Mekareeya:2017jgc}.

\begin{figure}[t]
	\centering
  	\includegraphics[scale=0.31]{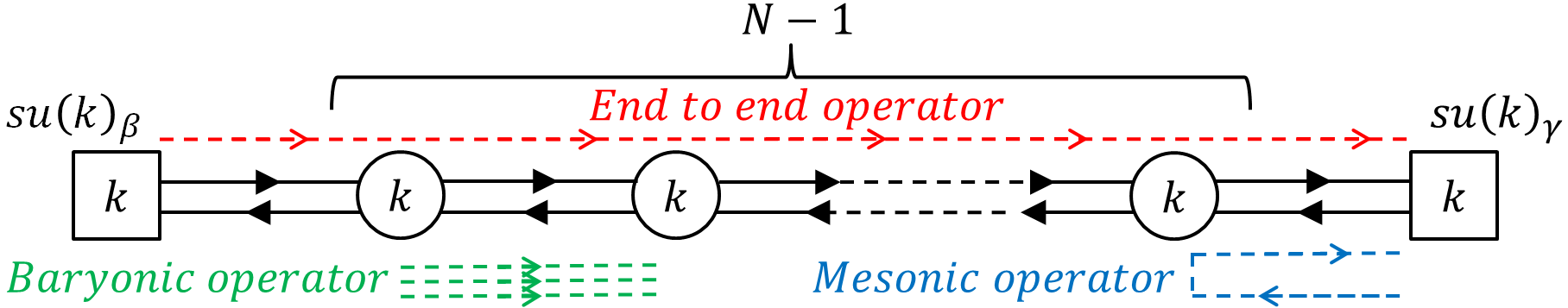}
    \caption{A quiver description of the theory $\mathcal{T}(SU(k),N)$ in six dimension on the tensor branch. Squares and circles represent $SU(k)$ flavor and gauge symmetries, respectively. The arrows represent half hypermultiplets in the bifundamental representation of the two adjacent symmetries. We present the three types of gauge invariant operators of the theory. In Red: We have the operators winding from one end of the quiver to the other. In blue: We have the mesonic operators winding from one end to the adjacent gauge node and back. In green: We have the baryonic operators, $k$ contractions of the same half hypermultiplet.}
    \label{F:6dquiver}
\end{figure}

Using the quiver description (see Fig. \ref{F:6dquiver}), we can look at the different gauge invariant operators we have. One important operator we will extensively discuss in this paper is the one winding from one end of the quiver to the other (see Fig. \ref{F:6dquiver} in red). This operator transforms in the fundamental (antifundamental) representation of $SU(k)_\beta$ and antifundamental (fundamental) representation of $SU(k)_\gamma$ and has charge $\pm N$ under $U(1)_t$ (here we have normalized $U(1)_t$ so that each bifundamental has charge $\pm1$). Another type of operator is the mesonic operator winding from one end to the neighboring gauge node and back to the same end (see Fig. \ref{F:6dquiver} in blue). These operators transform in the bifundamental representation (adjoint $\oplus$ singlet) of the same $SU(k)$ and have zero charge under $U(1)_t$. The last type of operators are the baryons (see Fig. \ref{F:6dquiver} in green), with charge $\pm k$ under $U(1)_t$ only. In the SCFT point there are only two such operators (with $\pm k$ $U(1)_t$ chrages), since all baryons with the same charge are identified with one another \cite{Hanany:2018vph}.

\begin{figure}[t]
	\centering
  	\includegraphics[scale=0.315]{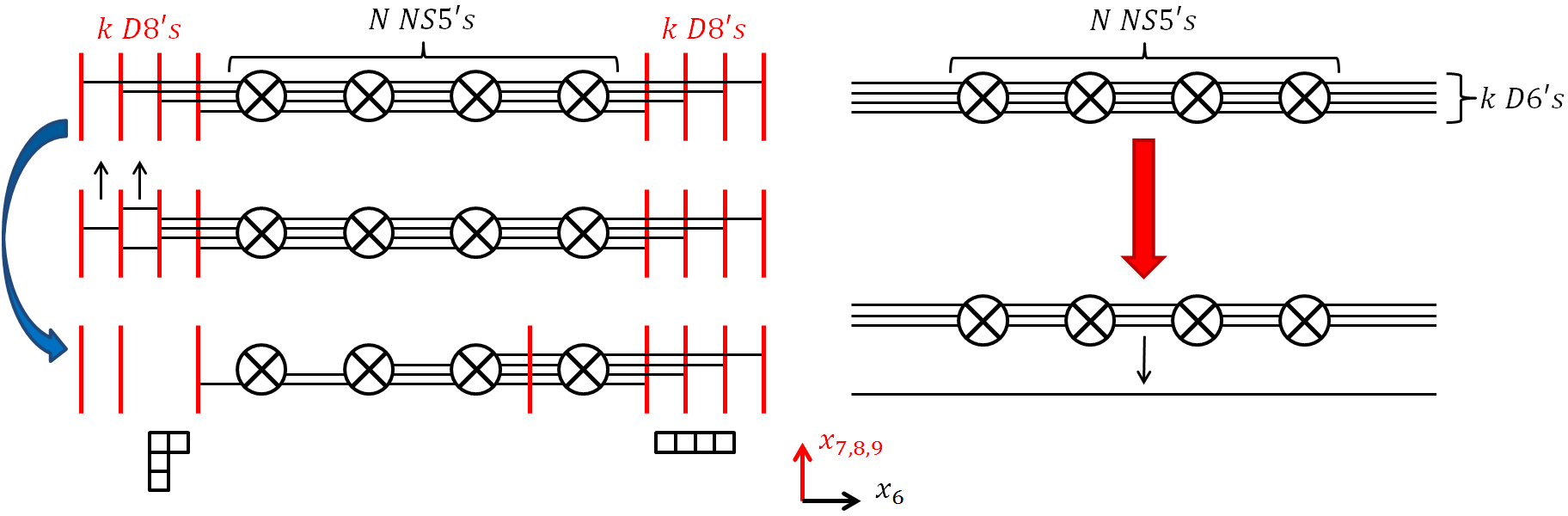}
    \caption{Type IIA brane constructions for theories of $\mathcal{T}(SU(k),N)$ (on the top of both sides) and their possible flows initiated by vevs to different operators. The $\otimes$ indicate NS5-branes occupying directions $0,...,5$, The black lines signify D6-branes in directions $0,...,6$, and the red lines denote D8-branes in directions $0,...,5,7,8,9$. \textbf{On the left:} The blue arrow signifies the flow generated by giving a vev to mesonic operators. In this case we first go on the partial Higgs branch sending several D6's to infinity. Finally we are left with several decoupled D8-branes and a quiver we can read from the brane picture after arranging branes using the Hanany-Witten effect. The flavor symmetry of the resulting theory can be denoted by two Young tableaux as shown bellow the brane construction. \textbf{On the right:} The red arrow symbols the flow initiated by giving a vev to the end to end operator. In this case we remain with a decoupled D6-brane and a quiver of $\mathcal{T}(SU(k-1),N)$. 
}
    \label{F:6dFlows}
\end{figure}

Another useful way to look at the $\mathcal{T}\left(SU(k),N\right)$ class of $6d$ theories, is by looking at the type IIA superstring theory brane construction of it arising from the reduction of the M-theory description we discussed before (see top of Fig. \ref{F:6dFlows}). This construction will prove very useful when giving vacuum expectation values (vev) to the above operators, since it allows to easily find the resulting IR theory at the end of the RG flow. We should also note that RG flows between SCFTs, including the particular class of theories we consider, were previously studied in \cite{HMRV,HR,HRT,HRT1}. We find that giving a vev to the end to end operators gets us to class $\mathcal{T}(SU(k'),N)$ with $k'<k$ (see Fig. \ref{F:6dFlows} red arrow). Giving a vev to the mesonic operator leads to class $\mathcal{T}(SU(k),\mu_L,\mu_R,N)$, where $\mu_L$ and $\mu_R$ specify a homomorphism $\mu_{L/R}:\mathfrak{su}(2)\rightarrow \mathfrak{su}(k)$ that can be represented by a Young tableaux (see Fig. \ref{F:6dFlows} blue arrow).\footnote{We ignore the process of giving a vev to the baryonic operator, as it is not relevant for this paper.} 

We will focus here on the compactifications of $\mathcal{T}(SU(k),N)$ models to four dimensions. We will only study flows triggered by vevs to operators winding  the quiver from end to end in the tensor branch description.
The resulting four dimensional models depend on several choices of the compactification, including, fluxes for $U(1)$ subgroups of the flavor symmetry, and choice of Riemann surface genus and punctures. When compactifying $\mathcal{T}(SU(k),N)$ models the resulting $4d$ class of theories is named class $\mathcal{S}_k$, and it has been studied comprehensively in the last few years \cite{Gaiotto:2015usa,Razamat:2016dpl,Bah:2017gph,Razamat:2018zus,Franco:2015jna,Hanany:2015pfa,Coman:2015bqq,Mitev:2017jqj,Bourton:2017pee}. The map between the choices of fluxes and surfaces in $6d$ and $4d$  theories in this class was discussed in \cite{Razamat:2016dpl,Bah:2017gph}. In any case of flux compactification from $6d$ models to $4d$, the flavor symmetry of the $6d$ theory will be partially broken in the $4d$ model according to the given flux.

\section{The tale of the two flows}
\label{S:TaleOf2Flows}
We start by considering the flow triggered by giving a vev to end to end operators. In $6d$ a natural choice is  to give a vev to one such operator, let us denote it ${\cal O}_{6d}$. We expect such a vev to generate a flow from $\mathcal{T}(SU(k),N)$ to $\mathcal{T}(SU(k-1),N)$. Compactifying the resulting theory on a Riemann surface with fluxes, $(\Sigma_{g',s'},{\cal F}')$, will lead to a $4d$ class $\mathcal{S}_{k-1}$ theory. We expect one can reach the same theory by first compactifying the $6d$ theory $\mathcal{T}(SU(k),N)$ on a Riemann surface with some fluxes, $(\Sigma_{g,s},{\cal F})$, to a class $\mathcal{S}_k$ theory, and then give a vev to some $4d$ operator which we denote by ${\cal O}_{4d}$. Given $(\Sigma_{g,s},{\cal F})$ and ${\cal O}_{4d}$ we want to understand what are the corresponding $(\Sigma_{g',s'},{\cal F}')$ and ${\cal O}_{6d}$.
The question of finding a dictionary between the two types of flows, $6d\to6d\to4d$ and $6d\to 4d\to 4d$, is a one of order of limits. A useful way to think about the problem is to start from $\mathcal{T}(SU(k),N)$ and turn on both deformation, the geometric one $(\Sigma_{g',s'},{\cal F'})$ and the vev for some ${\cal O}_{6d}$. The flow is then parametrized by the two scales, one set by the vev and another by the size of geometry. Taking one of these much larger than the other we should be able to derive $(\Sigma_{g,s},{\cal F})$ and ${\cal O}_{4d}$. 

The easier part is understanding what ${\cal O}_{4d}$ is. As the vev for ${\cal O}_{6d}$ breaks some symmetry of the six dimensional theory, a natural candidate for the corresponding vev in four dimensions is the operator ${\cal O}_{4d}$ which has exactly the same charges under all the symmetries as ${\cal O}_{6d}$. We will see in the following sections that in concrete cases it is very easy to find such a candidate. A less obvious question is how to determine $(\Sigma_{g,s},{\cal F})$ and that is what we will mainly do in this section. The complication comes from the fact that we turn on flux supported on the Riemann surface. 
What we will argue here, and later show in explicit field theory computations, is that the presence of flux has an effect of changing the number of minimal punctures\footnote{See Appendix \ref{A:PuncConventions} for class $\mathcal{S}_k$ puncture conventions.} between the two surfaces  $\Sigma_{g,s}$ and $\Sigma_{g',s'}$. That is $g=g'$ but $s\neq s'$. The reason for this is that the presence of flux forbids us from turning on a {\it constant} vev on the surface, instead the vev should be taken to depend on the position along the compactified directions. Let us discuss this issue in detail. 

We first consider the $6d\to6d$ part of the flow generated by a constant vev to ${\cal O}_{6d}$ in flat $6d$ space. From the low-energy tensor branch theory, this flow was triggered by vevs to scalar fields in the hypermultiplets, and is part of the Higgs branch of the initial $6d$ SCFT. In flat space this is a moduli of the theory, that is, it is a possible vacuum as the energy does not change. This is true in flat space and we next want to consider what happens when the $6d$ SCFT is compactified on a 2d surface with fluxes.

We immediately encounter the following problem. While a constant vev to a scalar in the hypermultiplet is a vacuum solution in flat space, it is generally not so once we compactify with fluxes, as the ordinary derivative is replaced with a covariant derivative and so a constant vev should change the kinetic energy.\footnote{The flow is ultimately performed at the $6d$ SCFT point, where the gauge theory description is inadequate. The flow is then better understood as a vev to a gauge invarint field. We are here using the tensor branch description as a way to understand the problem, though we expect the same issue to occur for the SCFT, as the gauge invariant may be charged under symmetries involving flux. It should be noted that the Higgs branch structure differs between the SCFT and the tensor branch gauge theory so some caution is advised here \cite{Hanany:2018vph}. This does not seem to effect the flow we consider here.} However, we may still hope that some modification of the vev, such that it will have some profile on the $2d$ surface, may still be a vacuum. We shall next explore this by restricting to the special case of genus one where we can map this problem to that of $5d$ theories connected by domain walls.

We shall first quickly review the approach used in \cite{Kim:2017toz,Kim:2018bpg,Kim:2018lfo} to tackle these types of compactifications. The starting point is to compactify the theory on one of the circles of the torus so that it flows to a $5d$ theory, potentially with a flavor holonomy. Without flux, it is known that for many cases, with a suitable choice of holonomy, the theory flows to a $5d$ gauge theory. This is just a generalization of the well known relation between the $(2,0)$ theory and $5d$ maximally supersymmetric Yang-Mills theory to less supersymmetric cases. Specifically for the cases we consider here, which can be thought of as $\mathbb{Z}_k$ orbifolds of the $A$ type $(2,0)$ theory; the $5d$ gauge theory is just a circular quiver of $k$ $SU(N)$ groups which is the known $\mathbb{Z}_k$ orbifold of an $SU(N)$ super Yang-Mills theory. In this case though, one also needs flavor holonomies breaking the $SU(k)^2$ global symmetry to its Cartan. We will not need any more details about the $5d$ gauge theory besides its existence.

\begin{figure}[t]
	\centering
  	\includegraphics[scale=0.3]{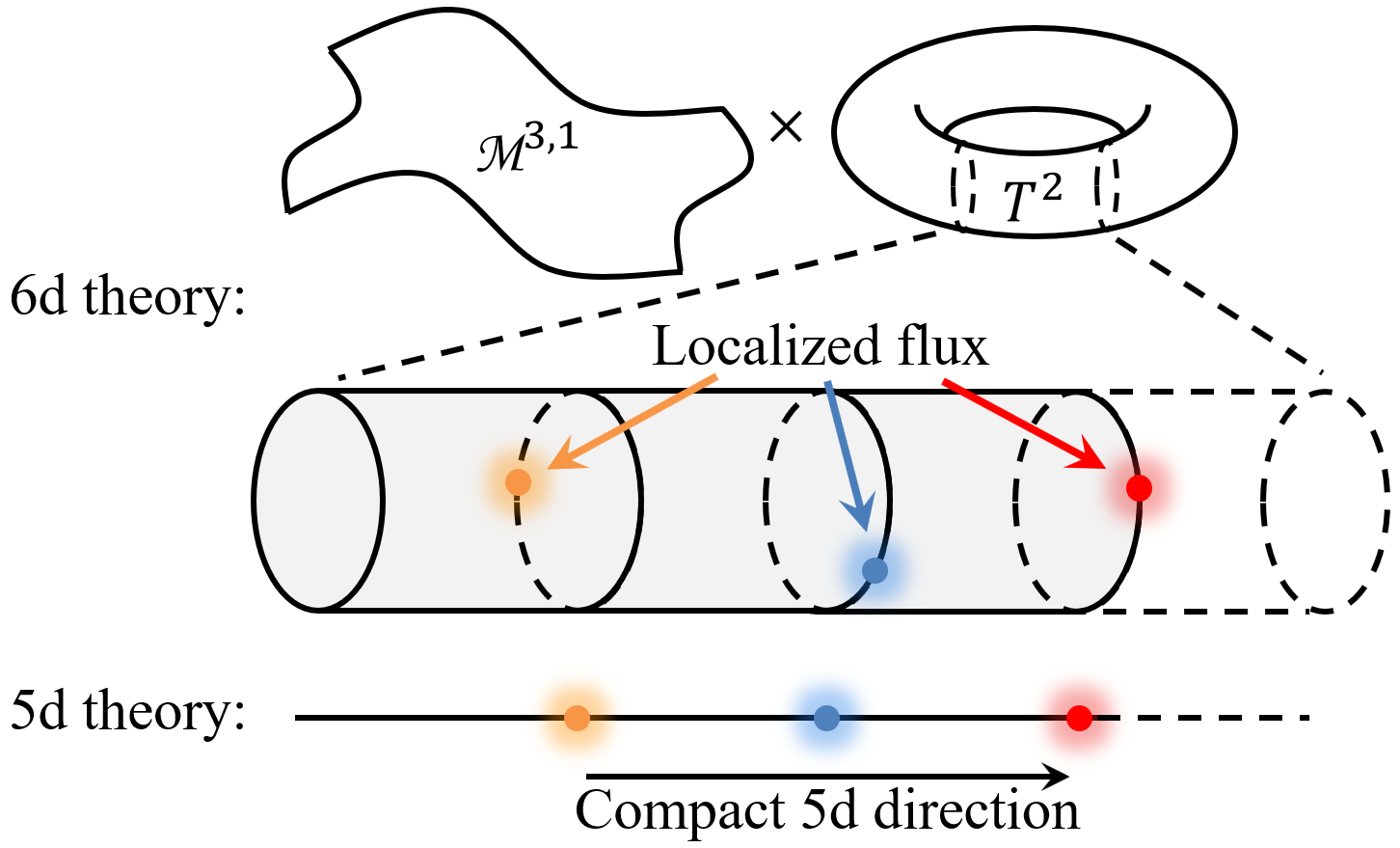}
    \caption{Top: An illustration of the $6d$ theory on a torus is shown in the upper picture. Beneath it we see a magnification of a cylinder out of the torus, where the circle represent the compactified direction taken to reduce the $6d$ theory to $5d$. Bottom: The compactification of the cylinder as part of the $5d$ theory. The compact $5d$ direction is the longitudinal direction. We denote in colors the localized flux in various locations along the compact $5d$ direction. These locations are described by $4d$ domain walls.}
    \label{F:5dLocalizedFlux}
\end{figure}

All of this is true when there is no flux, but next we want to consider the generalization once flux is included. We can take the flux into account by using an holonomy that varies along the compact $5d$ direction. Specifically we can consider an holonomy that has the profile of a sum of theta functions along the compact $5d$ direction, that is the other circle of the torus (see Fig. \ref{F:5dLocalizedFlux}). This should generate a flux in the form of a sum of delta functions along the compact $5d$ direction. In the regions where the flux is constant we expect to still get a $5d$ gauge theory depending on the chosen holonomy; however, as the holonomy changes, we have slightly different $5d$ gauge theories in different regions of the compact $5d$ direction. These regions connect to one another in points where the flux is located, and at these points there should be a $4d$ domain wall interpolating between the two theories (see Fig. \ref{F:5dLocalizedFlux}). Thus, the result of this analysis is that when considering the $5d$ compactification of theories with flux, the flux is manifested as domain walls between different $5d$ gauge theories that are low energy descriptions of the $6d$ SCFT compactified on a circle.\footnote{The compactified $6d$ SCFT generally only flows to a $5d$ gauge theory for special values of the holonomies. As a result this construction will only hold for cases where all the holonomies in the constant sections are of this type, and also only for fluxes that can be generated in this way. It is unclear if any flux can be realized using this construction. Nevertheless, for the cases we consider here, class $\mathcal{S}_k$ theories exhibiting the full $6d$ global symmetry, it is thought that all cases can be generated using this construction \cite{Kim:2018lfo}.}

Now let us consider the problem of giving a vev to a scalar field. As we previously mentioned in the presence of flux this is no longer a vacuum. However, we have now learned that we can think of this system as having the flux concentrated on codimension two defects in the $6d$ spacetime filling the four non-compact direction.\footnote{Here we take the connection to also have a delta function profile along the direction that we use to compactify from $6d$ to $5d$.} This implies that the field strength is zero almost everywhere on the Riemann surface, except at a finite number of points. At places where the field strength is zero, a scalar field profile that is equivalent to a constant up to a gauge transformation should be a vacuum, so the only issue is the behavior at the few points where the flux is located. It seems reasonable then that in the presence of flux we can still give a vev to a scalar that is mostly constant, up to a gauge transformation, with the exception of a few points with explicit dependence not removable by a gauge transformation. We would like to argue that it is this explicit dependence that leads to the appearance of minimal punctures. 

Before moving on to discuss the relation with minimal punctures, we first would like to address an issue. The previous discussion was rooted in a special choice of flux as concentrated on points on the $2d$ compactification surface. However, it is possible to choose more general fluxes without this property, so one may wonder if what we observed is an artifact of this choice. We would like to further consider this. First, we note that our construction share many similarities with vortex solutions in $3d$ gauge theories. In the simplest cases, the vortices are solutions of a system of scalar and vector fields were the vector field strength is concentrated in a finite region in space, while the scalar fields approach a constant, up to a gauge transformation, far away from that region, deviating from that as one goes in. We suspect that the configuration that we suggest, for more general fluxes, would just be a lift of these vortex solutions in $3d$, and so are not unusual in physics. The specific case we consider here is just the limit where the vortex size goes to zero. In supersymmetric theories this is expected to describe the massless vortices that are thought to play an important role in many superconformal $3d$ theories, and so in particular can also preserve  $4$ supercharges. 

We discussed till now flux configurations where the flux is concentrated on points on the $2d$ compactification surface.  One can wonder whether the specific form of the gauge field with a given value of flux can affect our conclusions. We expect that this is not the case and that the same conclusions hold if we smear the flux over the surface. A somewhat analogous claim is the fact that the $4d$ theories obtained in the compactifications from $6d$ depend only on complex structure moduli and not on the details of the metric on the Riemann surface, see {\it e.g} \cite{Anderson:2011cz}. On the other hand, we do expect that more than the total value of flux, other parameters associated with the connection may contribute in four dimensions. For example, it has been argued in many cases, see {\it e.g.}  \cite{Benini:2009mz,Bah:2012dg,Kim:2017toz,Razamat:2016dpl,Kim:2018bpg,Razamat:2018gro}, that flat connections reduce to exactly marginal couplings in four dimensions.\footnote{Although we do not discuss what would be the exact mechanism for this,  we do expect that the location of the localized flux on the surface can be mapped into the problem of analyzing the admissible flat connections. The basic idea behind this is that, in general, one can argue that dimensionally reducing conserved current operators in $6d$ one can obtain marginal operators in $4d$ and the problem of analyzing the dimensional reduction is tightly tied to the question of flat connections. See \cite{talknazareth,babuip} for details.} We do wish to note that the results we will show in the following section are qualitatively consistent with the expectations from this analysis, providing some confidence that the detailed distribution of the flux is not important for the expectations derived here.        

\begin{figure}[t]
	\centering
  	\includegraphics[scale=0.35]{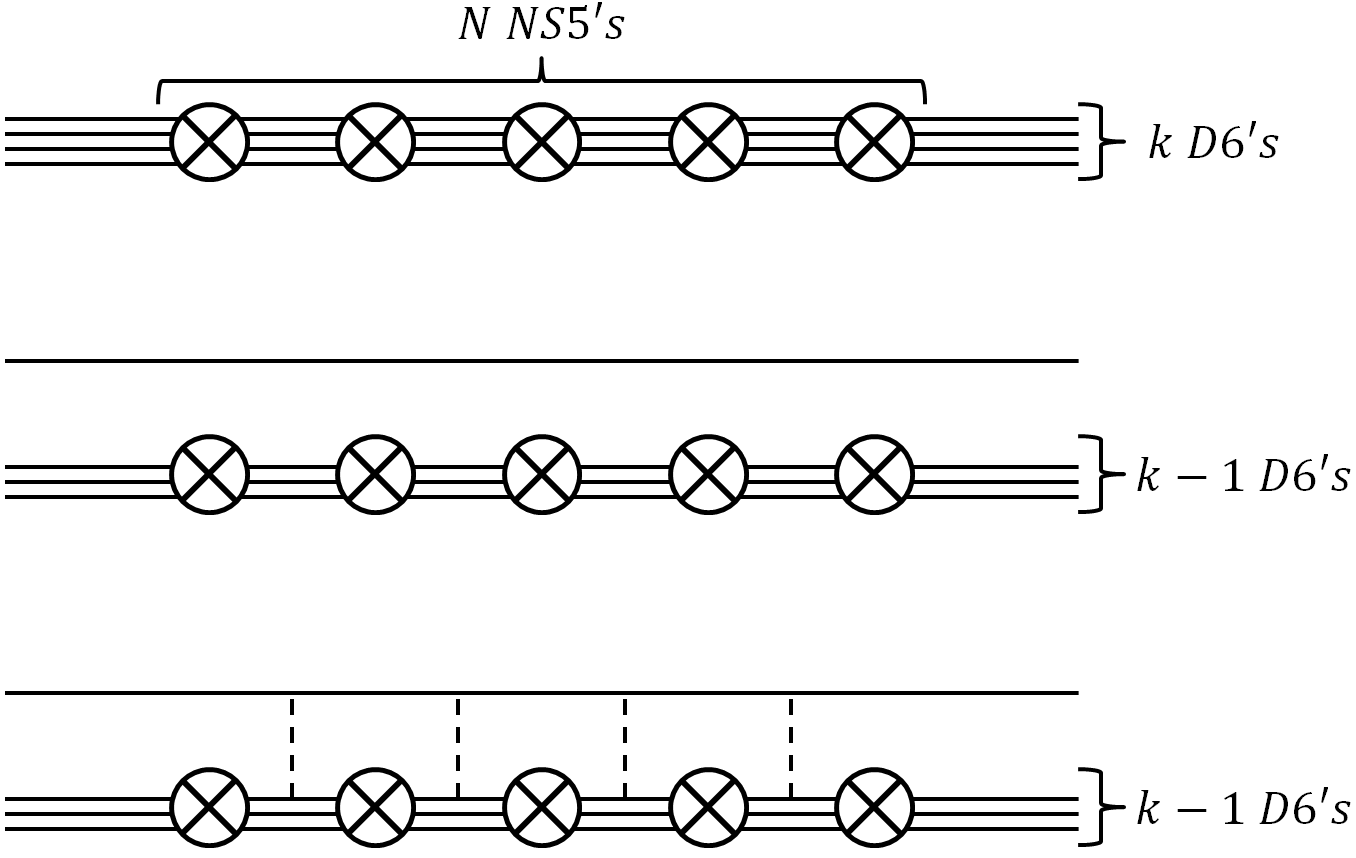}
    \caption{Type IIA brane construction. NS5-, D6- and D4-branes are represented by $\otimes$, solid lines and dashed lines, respectively, and fill directions 012345 (NS5's), 0123456 (D6's) and 01237 (D4's). Top: The brane construction matching the quiver in Fig. \ref{F:6dquiver} that can also be described by a stack of N M5-branes probing an $A_{k-1}$ singularity, and denoted by $\mathcal{T}(SU(k),N)$. Middle: A baryonic Higgs branch of the former theory flowing in the IR to $\mathcal{T}(SU(k-1),N)$. Bottom: We include additional D4-branes corresponding to a position dependent Higgs branch vev.}
    \label{F:SpaceDependentVEV}
\end{figure}

We now wish to return to the topic of why the flux leads to the appearance of minimal punctures. As we previously explained, we shall consider the parameter region, where the flux can be described as concentrated at points on the surface. We then expect that the vev should take the form of a covariant constant everywhere save at the location of the flux where it has some space dependence, leading to a codimension $2$ defect. In Fig. \ref{F:SpaceDependentVEV} we show the difference between a constant vev and a position dependent vev to the end to end operator on the tensor branch. This is shown as a brane construction similar to the one shown in Fig. \ref{F:6dFlows}. The difference is shown when one extracts one end to end D6 brane outside of the stack, and D4-branes stretch between the extracted D6-brane and the stack of remaining D6-branes.

These stretched D4-branes describe a vev with a spacetime dependence at special points \cite{Gaiotto:2014ina}. We again note that in brane systems describing supersymmetric $3d$ gauge theories, like the ones used in \cite{Hanany:1996ie}, the BPS vortices are described as D$1$ strings stretched between D$3$-branes \cite{Hanany:2003hp}. In these systems, the $3d$ gauge theory lives on the D$3$-branes, stretched between NS$5$-branes, and the ending D$1$ string has the effect of inserting magnetic flux to the D$3$-branes worldvolume theory, and so describes vortices. We also seek a brane description where one insert a flux on a codimension $2$ defect, and performing three T-dualities on the $3d$ brane system gives our proposed configuration. 

So we are lead to consider the brane configuration at the bottom of Figure \ref{F:SpaceDependentVEV} as the one appropriate for our case. We want to determine what is the effect of the additional D$4$-branes. Let us first consider the case where the end point of the flow is a theory in class $\mathcal{S}$. In that case we can first lift the configuration to M-theory, where it lifts to N M$5$-branes intersecting the additional M$5$-branes which are the lifts of the D$4$-branes. We next compactify this system on one of the circles of the torus, which is contained in the N M$5$-branes on which the $(2,0)$ theory lives, but is orthogonal to the other M$5$-branes. This should lead us to a system of N circular D$4$-branes intersected by NS$5$-branes. This is just the brane configuration describing the class $\mathcal{S}$ theory associated with a torus with minimal punctures described by the NS$5$-branes \cite{Gaiotto:2009we}. The addition of D$6$-branes just adds a $\mathbb{C}^2/\mathbb{Z}_k$ orbifold to the M-theory picture which then go to the same orbifold in the type IIA picture. The resulting picture is then the same with the additional NS$5$-branes mapping to minimal punctures. 

This leads us to conclude that when we give a vev to a $6d$ SCFT compactified on a torus with flux we expect to flow to a $4d$ theory associated with a compactification of the $6d$ SCFT at the end of the flow, but on a torus with additional minimal punctures. We expect the number of these minimal punctures to be proportional to the flux felt by the operator to which we are giving a vev, in particular, we expect no additional minimal punctures when the flux vanishes. This matches the formula for added minimal punctures we find in Section \ref{S:Coulomb limit}, which reduces for $g=1$ to
\be
m_{n}^{-}=c_{n}^{(k)}-b_{n}^{(k)} + N e_{k}\,.
\ee
In this formula, $m_n^-$ is the number of additional negative minimal punctures (see Appendix \ref{A:PuncConventions} for puncture definitions)  associated to the residual $U(1)_{\epsilon_n}$ symmetry that remains from the breaking of $U(1)_{\beta_n}\times U(1)_{\gamma_n}$ in  class $\mathcal{S}_k$. The vev we give breaks one combination of the $U(1)_{\beta_n}\times U(1)_{\gamma_n}$ and $U(1)_{\epsilon_n}$ is the other combination that remains unbroken. $b_n^{(k)}$, $c_n^{(k)}$ and $e_k$ are the fluxes under the class $\mathcal{S}_k$ symmetries $U(1)_{\beta_n}\times U(1)_{\gamma_n}\times U(1)_t$, respectively.\footnote{Recall, that when compactifying the $6d$ theory with general flux the global symmetry $SU(k)_\beta \times SU(k)_\gamma \times U(1)_t$ is generally broken to the maximal torus symmetry of $U(1)^{k-1}_\beta \times U(1)^{k-1}_\gamma \times U(1)_t$.}
Recall that here we are giving a vev to $4d$ operators corresponding to $6d$ operators charged in the bifundamental of $SU(k)_{\beta}\times SU(k)_{\gamma}$ and with charge $N$ under $U(1)_t$, and so the number of minimal punctures vanishes when the flux felt by the operator vanishes.

\begin{figure}[t]
	\centering
  	\includegraphics[scale=0.3]{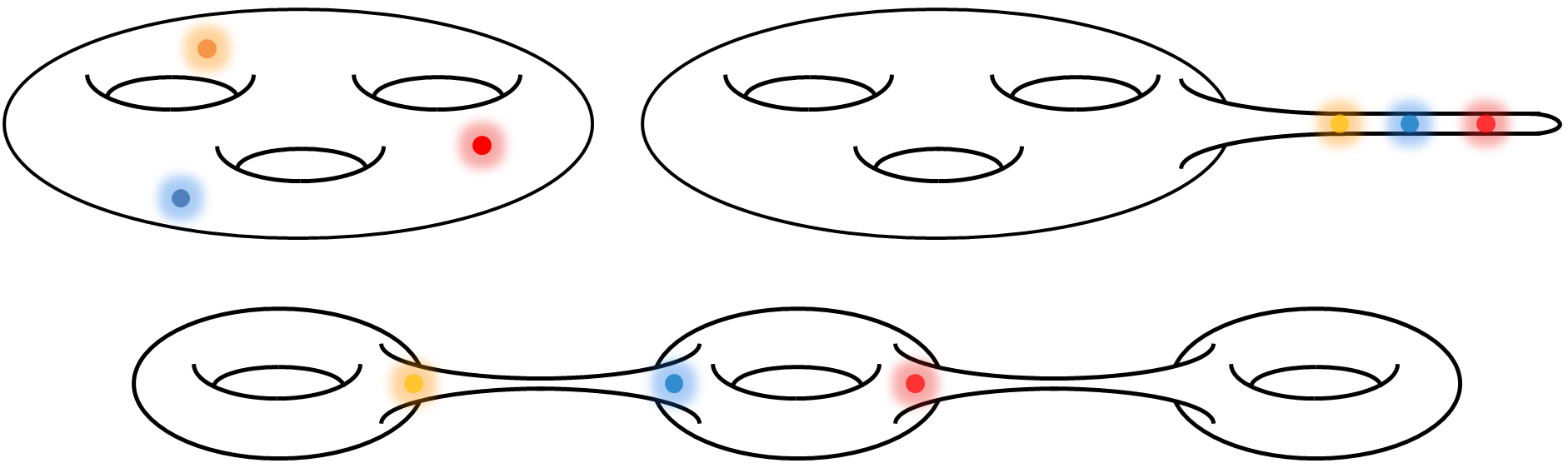}
    \caption{Illustrations of a general genus Riemann surface with localized fluxes denoted by colored points in different pair of pants decomposition schemes. \textbf{Top:} On the left a Riemann surface of a generic genus with some fluxes localized on it. On the right a pair of pants decomposition of the same surface where all the flux is along a long thin tube. \textbf{Bottom:} The same Riemann surface pair of pants decomposition as several tori connected by thin tubes. The metric is non flat only near the connection points of the tubes to the tori and the flux is chosen to be localized on the same points.}
    \label{F:LongFluxTube}
\end{figure}

We next want to consider the case of Riemann surfaces of generic genus. We suspect our previous argument for flavor symmetry fluxes to hold also in this case. The reason is that it is possible to take a pair of pants decomposition where all the flux is along a long thin tube connected to the rest of the surface (see top part of Fig. \ref{F:LongFluxTube}). We then expect the previous analysis to still apply to the long thin tube. Deforming to a specific pair of pants decomposition is again mapped to going to a special point on the conformal manifold, and there is the subtlety that the flow might be non-generic at that special limit, similarly to the previously discussed issue with the dependence on the exact flux configuration. Nevertheless, we shall see that the qualitative results we obtain are consistent with the results obtained from the Coulomb index in Section \ref{S:Coulomb limit}.

We now need to consider the effect of the curvature of the Riemann surface. Naively, as we are dealing with vevs to scalar fields, the curvature should not have any effect. However, the curvature leads to a breaking of supersymmetry unless we preform a twist, that is we couple the Cartan of the $6d$ $SU(2)_R$ symmetry to a background gauge field proportional to the curvature of the Riemann surface. This breaks the $SU(2)_R$ symmetry to its Cartan, which becomes an R-symmetry in $4d$ and shall be denoted as $U(1)^{6d}_R$. The scalar fields we give a vev to are charged under the $SU(2)_R$ symmetry and so are affected by the twist. As a result we expect the constant to again no longer be a solution because of the non-trivial R-symmetry connection. We can again tackle this problem in a similar manner to the flavor fluxes, as it is ultimately just a flux in a global symmetry. We again consider a pair of pants decomposition where the Riemann surface degenerate into a collection of tori connected via thin long tubes. In that case again the surface looks approximately like a flat surface except at the points where the tubes connect to the tori, and we can think of the flux as concentrated along these points (see bottom part of Fig. \ref{F:LongFluxTube}). As a result we expect the previous picture to still hold but now we need to also take into account the R-symmetry flux. We note that the full formula for added minimal punctures we derive in Section \ref{S:Coulomb limit} is given by
\be
m_{n}^{-}=c_{n}^{(k)}-b_{n}^{(k)} + N e_{k} + N(g-1)\,,
\ee 
and is consistent with this as the operator is the top component of an $\bold{N+1}$ dimensional $SU(2)_R$ multiplet and the twisting is such that it feels a flux of $N(g-1)$.

The rest of the changes then have the following interpretation. We expect the fluxes for the unbroken $\gamma$ and $\beta$ symmetries to change only though the effect of breaking the other $\gamma$ and $\beta$ symmetries, as these are not independent and are constrained to sum to zero. Indeed we will show this is consistent with the result
\be
b_{i}^{\left(k-\ell\right)} & = & b_{i+\ell}^{(k)}+\frac{1}{k-\ell}\sum_{n=1}^{\ell}b_{n}^{(k)}\qquad i=1,...,k-\ell\nonumber\\
c_{j}^{\left(k-\ell\right)} & = & c_{j+\ell}^{(k)}+\frac{1}{k-\ell}\sum_{n=1}^{\ell}c_{n}^{(k)}\qquad j=1,...,k-\ell\,,
\ee
derived in Section \ref{S:Coulomb limit}, where we consider a flow from class $\mathcal{S}_k$ to $\mathcal{S}_{k-\ell}$. The fluxes are just mapped to the previous one up to an overall shift, depending on the flux of the broken symmetries, that ensures that the sum of all the fluxes is zero.
    
The $U(1)_t$ and $U(1)^{6d}_R$ fluxes are a bit trickier, as the operator we are giving a vev to is charged under both of them. In general the vev should break the $U(1)$ that the operator is charged under and so the flux under it should be lost, which should lead to a change in flux. As a result, we expect that when the operator sees no flux there will also be no change in fluxes as we do not lose any flux. Indeed, in that case the number of additional punctures is zero and so the Riemann surface remains the same, which means that the $U(1)^{6d}_R$ flux remains unchanged. Also, we find the change of the $U(1)_t$ flux to be given by
\be
e_{k-\ell}=\frac{k}{k-\ell}e_{k}+\frac{\ell}{k-\ell}\left(g-1\right)-\frac{m_{tot}^{-}}{2\left(k-\ell\right)}\, ,
\ee
where $m_{tot}^{-}$ is the total number of additional minimal punctures. As a result when $m_{tot}^{-}$ vanishes the change in flux can be attributed entirely to the redefinition of $U(1)_t$ between the two $6d$ SCFTs expected from $6d$ analysis we give in Section \ref{S:6d flow}.   

Finally let us also mention that the flows will produce a variety of decoupled free fields in the IR which must be matched between the different flows. We will discuss this in the next section. 

\section{The $6d$ flow from $\mathcal{T}\left(SU(k),N\right)$ to $\mathcal{T}\left(SU(k-1),N\right)$}
\label{S:6d flow}
In this section we will consider the flow from $\mathcal{T}\left(SU(k),N\right)$ to $\mathcal{T}\left(SU(k-1),N\right)$ generated by giving a constant vev to the aforementioned end to end operators. 
Here we will first study the flow at the level of the anomaly polynomial. First, it is convenient to consider what we expect of this flow from field theory, where we shall employ the low-energy tensor branch gauge theory. In that description, this flow is generated via a vev to the gauge invariant made from the scalar fields in all the bifundamental hypermultiplets, including the ones at the edges, which are gauge-global symmetry bifundamentals. We recall here that $\mathcal{T}\left(SU(k),N\right)$ has an $SU(k)\times SU(k)\times U(1)_t$ global symmetry, in addition to the superconformal symmetry containing the $SU(2)_R$ R-symmetry. Under these symmetries, this gauge invariant is in the bifundamental of $SU(k)_\beta\times SU(k)_\gamma$, has charge $N$ under $U(1)_t$, and is in the $\bold{N+1}$ of $SU(2)_R$.     

We next wish to give a non-generic vev to this scalar field such that the $SU(k)_\beta\times SU(k)_\gamma\times U(1)$ global symmetry is broken to $SU(k-1)_\beta\times SU(k-1)_\gamma\times U(1)$. For that we first consider breaking each $SU(k)$ to $SU(k-1)\times U(1)$ such that
\be
\label{E:SUktoSUkm1}
\bold{k}_{SU(k)_\beta} \rightarrow \epsilon_\beta (\bold{k-1}_{SU(k)_\beta}) + \frac{1}{\epsilon_\beta^{k-1}} , \quad \bold{k}_{SU(k)_\gamma} \rightarrow \frac{1}{\epsilon_\gamma} (\bold{k-1}_{SU(k)_\gamma}) + \epsilon_\gamma^{k-1},
\ee
where we have introduced the fugacities $\epsilon_\beta$ and $\epsilon_\gamma$ for the $U(1)$ commutants of $SU(k-1)$ in $SU(k)$ for both $SU(k)_\beta$ and $SU(k)_\gamma$, respectively.\footnote{Here and throughout this paper we use fugacities to denote charges under $U(1)$ global symmetries. For example denoting an operator charges in terms of fugacities $a_1^{Q_1} a_2^{Q_2}\cdot ...\cdot a_N^{Q_N}$ means that the operator has charges $Q_i$ under $U(1)_{a_i}$, for $i=1,2,...,N$.}

We can next implement this decomposition on the gauge invariant that we built. We then find that there is a single $SU(k-1)\times SU(k-1)$ invariant in the decomposition, and we can think of this breaking as generated by a vev to this component. More specifically, as this field is in a non-trivial representation of $SU(2)_R$ we need to choose a specific component. This should break $SU(2)_R$ to its Cartan denoted by $U(1)^{6d}_R$, which will remain a global symmetry of the theory throughout the flow, though the superconformal symmetry is broken. As we are ultimately interested in supersymmetric theories also in $4d$, where $U(1)^{6d}_R$ is mapped to an R-symmetry, it is convenient to work with BPS components, which are usually those with highest R-charge. Therefore, we shall pick the component with $U(1)^{6d}_R$ charge $N$, where we have normalized $U(1)^{6d}_R$ such that the doublet has charges $\pm 1$. Thus, in field theory terms we are giving a vev to a scalar field with charges $\left(\epsilon_\gamma / \epsilon_\beta\right)^{k-1} t^{N} r^N$, where we use the fugacities $t$ and $r$ for $U(1)_t$ and $U(1)^{6d}_R$ respectively. This translates to setting
\be
\label{E:6dVEV}
U(1)_{\epsilon_\gamma} & = & -\frac{N}{k}U(1)_\epsilon - \frac{N}{2(k-1)}\left( U(1)^{6d}_R + U(1)_t \right) \,, \nonumber\\
U(1)_{\epsilon_\beta} & = & -\frac{N}{k}U(1)_\epsilon + \frac{N}{2(k-1)}\left( U(1)^{6d}_R + U(1)_t \right)\, .
\ee 

At the end of the flow we expect to get the $\mathcal{T}\left(SU(k-1),N\right)$ SCFT with several free hypers. This means that there must also be an $SU(2)_R$ at the end of the flow. However, its Cartan may not be just $U(1)^{6d}_R$ as it can potentially mix with other $U(1)$ symmetries. To determine the mapping we compare the charges of the scalar fields in the bifundamental hypermultiplets. Let us consider a single bifundamental between two adjacent gauge groups. In the $\mathcal{T}\left(SU(k),N\right)$ SCFT it is an $SU(2)_R$ doublet with charge $+1$ under $U(1)_t$ and in the bifundamental of the two adjacent $SU(k)$ gauge symmetries. We are giving a vev to a single component of each bifundamental, which causes all gauge groups to be Higgsed down to $SU(k-1)$. 

We can next decompose the two $SU(k)$ groups of the bifundamental to $SU(k-1)$ using similar decompositions as in \eqref{E:SUktoSUkm1}. The field we are giving a vev to is then charged as $(\epsilon_1 / \epsilon_2)^{k-1} t r$, where we use $\epsilon_1$ and $\epsilon_2$ for the fugacities of the $U(1)$ commutant of $SU(k-1)$ in $SU(k)$ for the two gauge groups. The vev then forces the identification $(\epsilon_2 / \epsilon_1)^{k-1} = t r$. Additionally there is also an $SU(k-1)\times SU(k-1)$ bifundamental in the decomposition. Its $U(1)$ charges are $\frac{\epsilon_2}{\epsilon_1} t(r + \frac{1}{r})$. After the identification forced by the vev, however, these become $t^{\frac{k}{k-1}} r^{\frac{1}{k-1}} (r + \frac{1}{r})$. This should map to the bifundamental in $\mathcal{T}\left(SU(k-1),N\right)$, which has charges $t' (r' + \frac{1}{r'})$, where we have used $t'$ and $r'$ for the fugacities of $U(1)_t$ and the Cartan of $SU(2)_R$ of $\mathcal{T}\left(SU(k-1),N\right)$. Matching the two we see that the symmetries are related as:
\be
\label{E:6dSymRel}
U(1)^{\mathcal{T}\left(SU(k-1),N\right)}_t & = & \frac{k}{k-1} U(1)^{\mathcal{T}\left(SU(k),N\right)}_t + \frac{1}{k-1} U(1)^{6d,\,\mathcal{T}\left(SU(k),N\right)}_R , \nonumber\\ U(1)^{6d,\,\mathcal{T}\left(SU(k-1),N\right)}_R & = & U(1)^{6d,\,\mathcal{T}\left(SU(k),N\right)}_R.
\ee  

With the above understandings we can set the vev and initiate the flow on the level of the $6d$ anomaly polynomial. The anomaly polynomial for $\mathcal{T}\left(SU(k),N\right)$ \cite{Ohmori:2014kda} is given by
\be
\label{E:AnomPol}
I_{8}^{\mathcal{T}\left(SU(k),N\right)} & = & \frac{(N-1)\left(k^{2}(N^{2}+N-1)+2\right)}{24}c_{2}^{2}(R)-\frac{(k^{2}-2)(N-1)}{48}c_{2}(R)p_{1}(T)\nonumber\\
 & & +\left(\frac{k}{24}p_{1}(T)-\frac{k(N-1)}{2}c_{2}(R)-\frac{kN}{2}c_{1}^{2}(t)\right)\left(c_{2}(\beta)_{\boldsymbol{k}}+c_{2}(\gamma)_{\boldsymbol{k}}\right)\nonumber\\
 & & -\frac{k}{6}\left(c_{4}(\beta)_{\boldsymbol{k}}+c_{4}(\gamma)_{\boldsymbol{k}}\right)+\left(\frac{k}{12}+\frac{1}{2}-\frac{1}{2N}\right)\left(c_{2}^{2}(\beta)_{\boldsymbol{k}}+c_{2}^{2}(\gamma)_{\boldsymbol{k}}\right)\nonumber\\
 & & +\frac{1}{N}c_{2}(\beta)_{\boldsymbol{k}}c_{2}(\gamma)_{\boldsymbol{k}}+\frac{k}{2}c_{1}(t)\left(c_{3}(\beta)_{\boldsymbol{k}}+c_{3}(\gamma)_{\boldsymbol{k}}\right)\nonumber\\
 & & +\frac{k^{2}N(N^{2}-1)}{12}c_{2}(R)c_{1}^{2}(t)-\frac{Nk^{2}}{48}p_{1}(T)c_{1}^{2}(t)+\frac{N^{3}k^{2}c_{1}^{4}(t)}{24}\nonumber\\
 & & +\frac{(7k^{2}+30N-30)}{5760}p_{1}^{2}(T)-\frac{(k^{2}+30N-30)}{1440}p_{2}(T) \, ,
\ee
where $c_2(R)$ is the second Chern class in the doublet of $SU(2)_R$ and $c_n(x)_{\boldsymbol{r}}$ is the $n$-th Chern class in the representation $\boldsymbol{r}$ of the symmetry associated to $x$. $\beta$, $\gamma$ and $t$ are associated to the symmetries $SU(k)_\beta$, $SU(k)_\gamma$ and $U(1)_t$, respectively. $p_1(T)$ and $p_2(T)$ are the first and second Pontryagin classes of the tangent bundle.

We next need to decompose the $SU(k)$ symmetries to $SU(k-1)$ using \eqref{E:SUktoSUkm1}. This is implemented in the anomaly polynomial by the assignments
\be
c_2 (SU(k)_\beta)_{\bold{k}} & = & c_2 (SU(k-1)_\beta)_{\bold{k-1}} - \frac{k(k-1)}{2} c^2_1 (U(1)_{\epsilon_\beta}) ,\nonumber\\
c_3 (SU(k)_\beta)_{\bold{k}} & = & c_3 (SU(k-1)_\beta)_{\bold{k-1}} - 2 c_1 (U(1)_{\epsilon_\beta})c_2 (SU(k-1)_\beta)_{\bold{k-1}}\nonumber\\
 & & - \frac{k(k-1)(k-2)}{3} c^3_1 (U(1)_{\epsilon_\beta}) ,\nonumber\\
c_4 (SU(k)_\beta)_{\bold{k}} & = & c_4 (SU(k-1)_\beta)_{\bold{k-1}} - \frac{(k^2 - k- 6)}{2} c^2_1 (U(1)_{\epsilon_\beta}) c_2 (SU(k-1)_\beta)_{\bold{k-1}} \nonumber \\ 
& & -3 c_1 (U(1)_{\epsilon_\beta})c_3 (SU(k-1)_\beta)_{\bold{k-1}} - \frac{k(k-1)(k-2)(k-3)}{8} c^4_1 (U(1)_{\epsilon_\beta})\, ,\nonumber\\
\ee
where $c_n(SU(k)_\beta)\equiv c_n(\beta)$. There are similar assignments under the exchange $\beta\rightarrow \gamma$, but with $c_1 (U(1)_{\epsilon_\beta})\rightarrow -c_1 (U(1)_{\epsilon_\gamma})$ due to the different definitions of the symmetries in \eqref{E:SUktoSUkm1}. In addition, we will also break $SU(2)_R$ to its Cartan by taking $c_2 (R) = - c_1 (R')^2$. Finally, we will initiate the flow by giving the aforementioned vev translating to the assignments in \eqref{E:6dVEV}. Under all these assignments \eqref{E:AnomPol} transforms to
\be
I_{8}^{\mathcal{T}\left(SU(k),N\right),Flow} & = & I_{8}^{\mathcal{T}\left(SU(k-1),N\right)}\left(c_1(t)\rightarrow \frac{k}{k-1}c_1(t)+\frac{1}{k-1} c_1(R'),\,c_2 (R) \rightarrow - c_1 (R')^2\right)\nonumber\\ 
 & & + I_8^{\text{free hyper}} + I_8^{\beta \text{ free hypers}} + I_8^{\gamma \text{ free hypers}}\, ,
\ee
where the last three terms have the form of an anomaly polynomial of free hypermultiplets. Specifically, $I^{\text{free hyper}}_8$ can be identified with the anomaly polynomial of a free half-hypermultiplet in the doublet of $SU(2)_R$. $I^{\beta \text{ free hypers}}_8$ can be identified with the anomaly polynomial of $k-1$ free hypers in the fundamental of $SU(k-1)_\beta$, charge $k$ under $U(1)_{\epsilon_\beta}$ and charge $-1$ under $U(1)^{6d}_R$. Similarly, $I^{\gamma \text{ free hypers}}_8$ can be identified with the anomaly polynomial of $k-1$ free hypers in the fundamental of $SU(k-1)_\gamma$, charge $-k$ under $U(1)_{\epsilon_\gamma}$ and charge $-1$ under $U(1)^{6d}_R$. These charges will be trivially shifted by the identifications of \eqref{E:6dVEV}.

From the above result we expect the matching decoupled free chiral multiplets in $4d$ to be with the following charges\footnote{We chose the free chirals to be in the anti-fundamental of $SU(k-1)_\beta$ and the fundamental of $SU(k-1)_\gamma$, as this choice matches the results we will find in Section \ref{S:Coulomb limit} for $4d$. This stems from the identification of the $6d$ SCFT symmetries, as defined here, and the conventions of class $\mathcal{S}_k$, as set out in \cite{Gaiotto:2015usa}.}
\be
FC_\beta & = & \sum_{i=1}^{k-1}\beta_i^{-N}\epsilon_\beta^k = \sum_{i=1}^{k-1}\beta_i^{-N}\epsilon^{-N} \left(\frac{r}{t}\right)^{\frac{Nk}{2(k-1)}}\,, \nonumber\\
FC_\gamma & = & \sum_{j=1}^{k-1}\gamma_j^{N}\epsilon_\gamma^{-k} = \sum_{i=1}^{k-1}\gamma_j^{N}\epsilon^{N} \left(\frac{r}{t}\right)^{\frac{Nk}{2(k-1)}}\,,
\ee
where we used the definition $U(1)_t^{4d}=-U(1)_t^{6d}$. In addition, $\beta_i$ and $\gamma_j$ are the Cartan of $SU(k-1)_\beta$ and $SU(k-1)_\gamma$, respectively, as we break them when we compactify to $4d$ and give flux to these global symmetry. These Cartan charges uphold the relations $\prod_{i=1}^{k-1} \beta_i=1$ and $\prod_{j=1}^{k-1} \gamma_j=1$. The second equality signs are due to the relations in \eqref{E:6dVEV}. We would like to write the charges in terms of the $\mathcal{T}\left(SU(k-1),N\right)$ theory $t$- and $R$- charges, and also move to the R-charge conventions used in $4d$ as $t\rightarrow t-r_{4d}$ and $r_{6d}\rightarrow r_{4d}$. We find that the expected free chiral fields take the form
\be
FC_\beta & = & \sum_{i=1}^{k-1}\beta_i^{-N}\epsilon^{-N} \left(\frac{2r}{t}\right)^{N/2}\,, \nonumber\\
FC_\gamma & = & \sum_{i=1}^{k-1}\gamma_j^{N}\epsilon^{N} \left(\frac{2r}{t}\right)^{N/2}\,.
\ee

Using this flow from $\mathcal{T}\left(SU(k),N\right)$ to $\mathcal{T}\left(SU(k-1),N\right)$, one can easily generalize to flow to $\mathcal{T}\left(SU(k-\ell),N\right)$. For instance equation \eqref{E:6dSymRel} can be generalized to that case giving:
\be
\label{E:6dSymRelGen}
U(1)^{\mathcal{T}\left(SU(k-\ell),N\right)}_t & = & \frac{k}{k-\ell} U(1)^{\mathcal{T}\left(SU(k),N\right)}_t + \frac{\ell}{k-\ell} U(1)^{\mathcal{T}\left(SU(k),N\right)}_R , \nonumber\\ U(1)^{\mathcal{T}\left(SU(k-\ell),N\right)}_R & = & U(1)^{\mathcal{T}\left(SU(k),N\right)}_R.
\ee
This is then the relation between these symmetries for the two theories. As for the expected charges of the decoupled free fields in $4d$ we find it generalizes to
\be
\label{E:FC6d}
FC_{\beta,n} & = & \sum_{i=1}^{k-\ell}\beta_i^{-N}\epsilon_n^{-N} \left(\frac{2r}{t}\right)^{N/2}\,, \nonumber\\
FC_{\gamma,n} & = & \sum_{i=1}^{k-\ell}\gamma_j^{N}\epsilon_n^{N} \left(\frac{2r}{t}\right)^{N/2}\,.
\ee
where $\prod_{i=1}^{k-\ell} \beta_i=1$ and $\prod_{j=1}^{k-\ell} \gamma_j=1$. In addition we may expect in this general case free chirals charged under two of the $\epsilon_n$-s only, related to strings attached to two of the $D6$-branes removed from the stack in the brane picture.

\section{The Coulomb branch limit}
\label{S:Coulomb limit}

\begin{figure}[t]
	\centering
  	\includegraphics[scale=0.22]{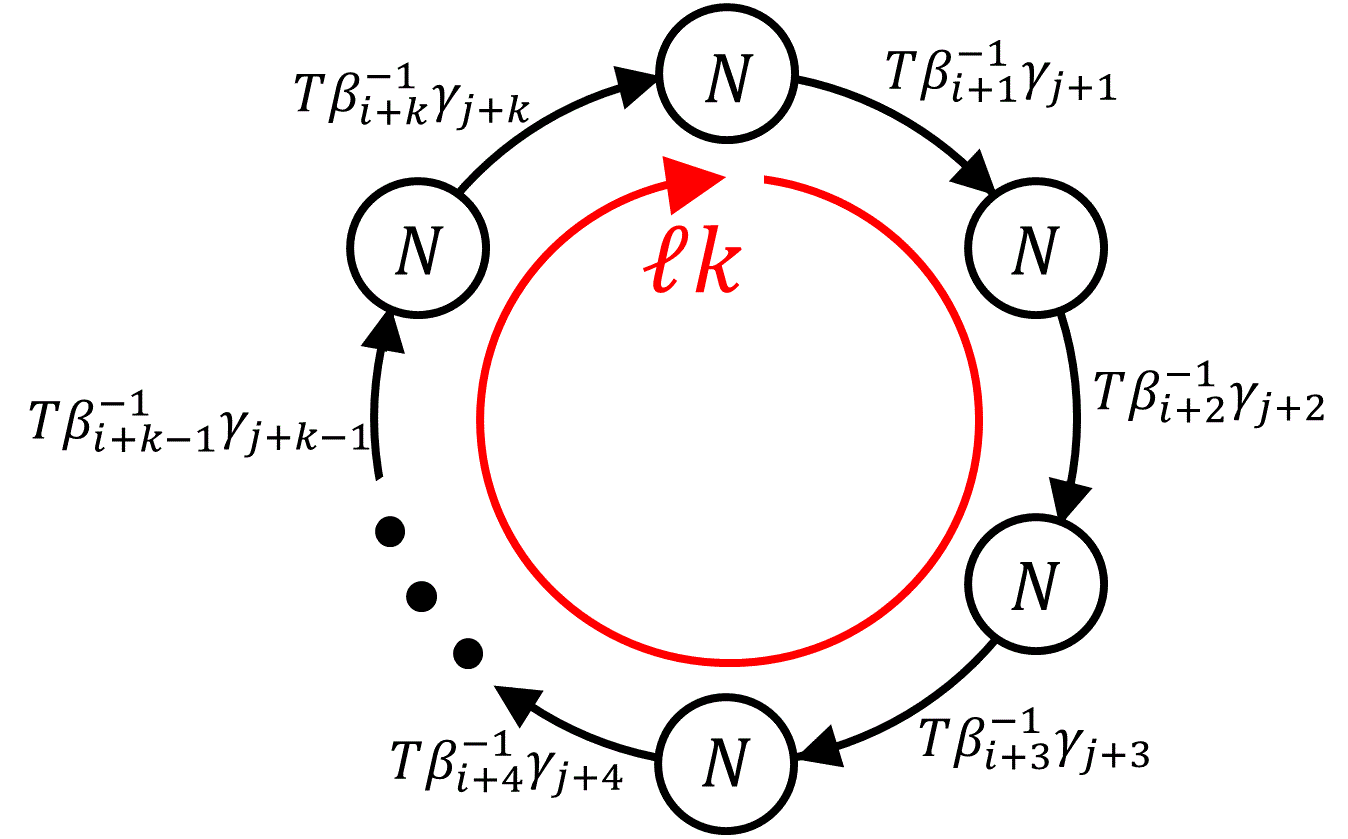}
    \caption{The fields added in $\Phi$ gluing. The baryonic operators $T^N \beta_i^{-N} \gamma_j^N$ introduced in the gluing are the ones we give vacuum expectation value to. Not all operators exist in every model, and the spectrum depends on the fluxes (and also puncture properties in case there are ones). We therefore expect the flows to depend non trivially on the fluxes. $T$ indicates an R-charge 2 and $U(1)_t$ charge $-1$ (this is related to the $U(1)_t$ symmetry originating from the symmetries of the $6d$ theory).  }
    \label{F:ChiralRing}
\end{figure}

Armed with the knowledge of known $6d$ operators and the resulting flows generated from giving vevs to these operators; we can try to map them to the matching $4d$ theories and operator vevs. These map to $4d$ baryonic operators that exist in some of the $4d$ theories depending on the Riemann surface and fluxes.\footnote{In the case of a punctured Riemann surface it also depends on the puncture properties.} These operators are explicitly added in the process of gluing two punctured Riemann surfaces by the so called $\Phi$-gluing, see Appendix \ref{A:PuncConventions} for puncture and gluing types. The aforementioned operators are illustrated in Fig. \ref{F:ChiralRing}.

The operators we give a vev to are captured by a specific limit of the superconformal index of class $\mathcal{S}_k$ theories \cite{Razamat:2018zus}, which generalizes the Coulomb index of class ${\cal S}$ models \cite{Gadde:2011uv}. 
For $\mathcal{N}=2$ models the Coulomb limit only captures operators from the Coulomb branch, which for Lagrangian theories are  of the form $Tr\phi^k$ with $\phi$ being the adjoint chiral in the vector multiplet.  Specifically, for class $\mathcal{S}$ such operators are added when gluing two punctures as they belong to the ${\cal N}=2$ vector multiplet. For $\mathcal{N}=1$ the notion of the Coulomb branch is not as natural as the one for ${\cal N}=2$, but for the specific case of class $\mathcal{S}_k$ it was found in \cite{Razamat:2018zus} that there is a limit of the superconformal index that counts operators that are added when gluing two punctures, see Figure \ref{F:ChiralRing}. This limit of the index is obtained by taking the limit of $p,q,t\to 0$ while keeping $T\equiv\frac{pq}{t}$ constant, where $t$ is the fugacity associated with the $U(1)_t$ symmetry coming from $6d$, and $p$ and $q$ are superconformal fugacities defined in Appendix \ref{A:indexdefinitions}. Additionally for $k=1$ (no singularity) this limit is the same as the class $\mathcal{S}$ Coulomb limit.

The operators we are interested to give a vev to, are some of the operators contributing to the Coulomb limit described above (see Fig. \ref{F:ChiralRing}). Therefore, we can use the Coulomb limit formula as a simple tool to help us map the flows from $\mathcal{S}_k$ to class $\mathcal{S}_{k'}$ with $k'<k$, and to identify the new fluxes. Additionally, we can use the explicit Lagrangians known in class $\mathcal{S}_k$ to find on the level of the superconformal index the mapping of the flows in some specific examples. We will also initiate the flow on the known class $\mathcal{S}_k$ anomaly polynomial to verify further our results.

We will concentrate on deriving results and mapping the flows from the Coulomb index in this section. To that end we start by recalling the result found in \cite{Razamat:2018zus}, the Coulomb limit formula for class $\mathcal{S}_k$ on a Riemann surface without punctures was given as
\be
\label{E:Coulomb}
\mathcal{I}_{g,\left(b_{i},c_{j},e\right)}^{N,k} & = & PE\left[\sum_{i,j=1}^{k}\left(-b_{i}+c_{j}+N e+\left(N-1\right)\left(g-1\right)\right)\beta_{i}^{-N}\gamma_{j}^{N}T^{N}\right]\times\nonumber\\
& & PE\left[\sum_{\ell=1}^{N-1}\left(\ell k e+\left(\ell k -1\right)\left(g-1\right)\right)T^{\ell k}\right] \, ,
\ee 
with $N$ the number of probing M5-branes, $k$ related to the $A_{k-1}$ singularity and $g$ the genus of the Riemann surface. Here each $T\equiv \frac{pq}{t}$ adds $2$ to the R-charge and $-1$ to the $U(1)_t$ charge, and $\left(b_{i},c_{j},e\right)$ are, as defined before, the fluxes for $\left(\beta_{i},\gamma_{j},t\right)$, the residual (Cartan subalgebra) $U(1)$ internal symmetries remaining from the $6d$ $SU(k)_\beta \times SU(k)_\gamma \times U(1)_t$ flavor symmetry. $PE[...]$ is the plethystic exponent, see Appendix \ref{A:indexdefinitions} for definition. We will mostly deal with closed Riemann surfaces not including punctures to keep the main results simple and not clutter them with the large amount of obstructions that punctures create. Some comments and results with punctures are available in Appendix \ref{A:Flows punc}.

When we give a vacuum expectation value to an operator we set its scalar component to some value that relates to some energy scale. This value is not charged under any global symmetry (including R-symmetry), forcing to identify some of the symmetries that the operator is charged under with one another. In the superconformal index language the symmetries appear as fugacities which are charges under the Cartan subalgebra of each symmetry. When we give a vev to some operator we simply set the combination of fugacities it is charged under to be $1$. For example in what follows we will give vevs to operators with charge $-N$ under one of the $U(1)_{\beta_{i}}$ symmetries, charge $N$ under one of the $U(1)_{\gamma_{j}}$ symmetries, charge $-N$ under $U(1)_t$ and R-charge $2N$ under the conventional R-charge used in the Coulomb limit formula.\footnote{This is not necessarily the conformal R-charge.} In fugacities this vev translates to setting $\left(\beta_{i}^{-1}\gamma_{j}\left(\frac{pq}{t}\right)\right)^{N}=1$.

The general formula for the Coulomb limit of the superconformal index  is valid for all theories of class $\mathcal{S}_k$ (modulo some restrictions on the flux, see \cite{Razamat:2018zus}). Some of these, at the moment, lack a simple Lagrangian definition,  whereas  for others such a description is known. For theories with no known Lagrangian what physically happens during the flow is hard to analyze, whereas for Lagrangian models we can easily follow the flow.
Thus, we can only understand the flow process for Lagrangian theories that have the required operators, and we give some account for many such examples in Section \ref{S:IndexComp}.
In some of these Lagrangian cases we generate the flow by triggering vevs to baryonic operators composed of fields with charges $\beta_{i}^{-1}\gamma_{j}\left(\frac{pq}{t}\right)$, which are  the ones added in $\Phi$-gluing, see Figure \ref{F:ChiralRing}.
Considering any specific theory, including non-Lagrangian ones, one can easily see in the formula for the Coulomb index that the required operators exist by checking if the coefficient of $\beta_{i}^{-N}\gamma_{j}^N T^N$ in the plethystic exponent is greater than zero. The flow triggered by the aforementioned vevs then may include Higgsing of some of the gauge symmetries and also fields becoming massive depending on the specifics of the UV theory.

In this section we will first map flows that break all the $\beta_i$ and $\gamma_j$ internal symmetries, as these flows leave us with a simple result in class $\mathcal{S}$ allowing an easy mapping of fluxes and free fields. After this, we will consider vevs partially breaking the $\beta_i$ and $\gamma_j$ internal symmetries, resulting in class $\mathcal{S}_{k-\ell}$ with no additional punctures. As before we will find the mapping of fluxes and free fields for this case. Finally, we will consider vevs breaking some of the pairs $U(1)_{\beta_i}\times U(1)_{\gamma_i}$ to a residual $U(1)_{\epsilon_i}$. As we discussed before, in the general case this will result in class $\mathcal{S}_{k-\ell}$ theories with additional minimal punctures. We will find the number of additional punctures and map fluxes and free fields in this case as well. We will show that the result of the flow with additional minimal punctures can be related to the former result without minimal punctures by simply closing the minimal punctures. This sequence of results should allow to track our logic and mappings as we move from the simplest result with a minimal number of mapping parameters to the most complicated one.

\subsection{Flows from class $\mathcal{S}_k$ to class $\mathcal{S}_1$}
\label{SS:Coulomb Flow WO Punct N=k=2}
Starting from the simplest operator vevs, we first wish to break all internal symmetries excluding $U(1)_t$ ($\beta$'s and $\gamma$'s only). This should lead us to class $\mathcal{S}_1$ which is the same as class $\mathcal{S}$ up to free fields.\footnote{We use class $\mathcal{S}_1$ and not class $\mathcal{S}$ in order to make the generalization of the results more obvious later on, when we discuss flows from class $\mathcal{S}_k$ to $\mathcal{S}_{k'}$ with $k>k'$.} 
The triggering of the vevs is implemented in the index by setting $\left(\beta_{i}^{-1}\gamma_{j}\left(\frac{pq}{t}\right)\right)^{N}=1$ with $i=1,...,k-1$ and $j=1,...,k$. These vevs translate to the assignments $\beta_{i}=\frac{pq}{t},\gamma_{j}=1$. 
Placing the above assignment in the Coulomb limit appearing in \eqref{E:Coulomb} we find,
\be
\label{E:CoulombFlow}
\mathcal{I}_{g,\left(b_{i},c_{j},e_k\right)}^{N,k(flow)} & = & PE\left[\left(kb_{k}+N k (k-1)  e_k+\left(N-1\right)\left(g-1\right) k (k-1) \right)\right]\times\nonumber\\
 & &  PE\left[\left(-k b_{k}+N k e_k+\left(N-1\right)\left(g-1\right) k \right) T^{k N}\right]\times\nonumber\\
 & & PE\left[\sum_{\ell=1}^{N-1}\left(\ell k e_k+\left(\ell k -1\right)\left(g-1\right)\right)T^{\ell k}\right] \, .
\ee

We next want to compare the above formula to the Coulomb limit formula of class $\mathcal{S}_1$, but first we want to shift the charges under the $U(1)_t$ to match the expected powers of $T$ contributing to the limit in class $\mathcal{S}_1$. This is achieved by redefining charges such that $T\rightarrow T^{1/k}$. The limit transforms accordingly to
\be
\mathcal{I}_{g,\left(b_{i},c_{j},e_k\right)}^{N,k(flow)} & = & PE\left[\left(-k b_{k}+N k e_k+\left(N-1\right)\left(g-1\right) k \right) T^{N}\right]\times\nonumber\\
 & & PE\left[\sum_{\ell=1}^{N-1}\left(\ell k e_k+\left(\ell k -1\right)\left(g-1\right)\right)T^{\ell}\right] \, .
\ee   
Here we stripped the divergence that appeared in the first line. Comparing to the Coulomb limit of class $\mathcal{S}_1$
\be
\mathcal{I}_{g,\left(e_1\right)}^{N,k=1} & = & PE\left[\left(N e_1+\left(N-1\right)\left(g-1\right)\right)T^{N}\right]\times\nonumber\\
& & PE\left[\sum_{\ell=1}^{N-1}\left(\ell e_1+\left(\ell -1\right)\left(g-1\right)\right)T^{\ell}\right] \, .
\ee
We can match all coefficients of $T^\ell$ with $\ell=1,...,N-1$ by setting the flux of the $\mathcal{S}_1$ theory to be
\be
\label{E:tFluxFlow}
e_1 = k e_k +\left(k-1\right) \left(g-1\right) .
\ee
The coefficients of $T^N$ don't match for this value, but we will show that this mismatch corresponds to free chiral multiplets of R-charge $2N$ and $U(1)_t$ charge $-N$ that decouple in the flow. The comparison allows to find the number of decoupled chiral multiplets, given by
\be
\label{E:freeChirals}
n_{FC}=-kb_{k}-\left(k-1\right)\left(g-1\right)\,.
\ee
In case the number of decoupled free chirals is negative it means in the simplest sense, that one needs to decouple fields with opposite charges and $(2-Q_R)$ R-charge, where in this case $Q_R=2N$ is the original R-charge of the free chiral fields.

Other vev choices that break all the internal symmetries (except $U(1)_t$), can be easily mapped to this specific choice by replacing $b_k$ with either $b_i$ or $c_j$ matching the vevs.

\subsection{Flows from class $\mathcal{S}_k$ to class $\mathcal{S}_{k-\ell}$}
\label{SS:Coulomb Flow WO Punct}

After understanding the Coulomb limit flow followed by the simple vevs breaking all the $\beta$ and $\gamma$ internal symmetries, we wish to understand the flow generated by vevs that partially break the internal symmetries. For this purpose we choose baryon vevs $\left(\beta_{n}^{-1}\gamma_{m}\left(\frac{pq}{t}\right)\right)^{N}=1$ and $\left(\beta_{n}^{-1}\gamma_{k}\left(\frac{pq}{t}\right)\right)^{N}=1$, meaning we set $\beta_{n}=\frac{pq}{t}\gamma_{k},\gamma_{m}=\gamma_{k}$, with $n,m=1,..,\ell$. In addition we make the identification of the new internal symmetries such that $\beta_{i}=\widetilde{\beta}_{i-\ell}\left(\frac{pq}{t}\right)^{-\ell/\left(k-\ell\right)}\gamma_{k}^{-\ell/\left(k-\ell\right)}$ and $\gamma_{j}=\widetilde{\gamma}_{j-\ell}\gamma_{k}^{-\ell/\left(k-\ell\right)}$, with $i,j=\ell+1,...,k$. Specifically, the assignments required are
\be
\label{E:General vevs}
\beta_{n}=\frac{pq}{t}\widetilde{\gamma}_{k-\ell}^{\left(k-\ell\right)/k}\,,&\;\beta_{i}=\widetilde{\beta}_{i-\ell}\left(\frac{pq}{t}\right)^{-\ell/\left(k-\ell\right)}\widetilde{\gamma}_{k-\ell}^{-\ell/k}&\qquad n=1,...,\ell,\quad i=\ell+1,...,k\nonumber\\
\gamma_{m}=\widetilde{\gamma}_{k-\ell}^{\left(k-\ell\right)/k}\,,&\;\gamma_{j}=\widetilde{\gamma}_{j-\ell}\widetilde{\gamma}_{k-\ell}^{-\ell/k}&\qquad m=1,...,\ell,\quad j=\ell+1,...,k\,.
\ee
Using these assignments in the Coulomb limit \eqref{E:Coulomb}, it transforms to
\be
\mathcal{I}_{g,\left(b_{i},c_{j},e\right)}^{N,k(flow)} & = & PE\left[\ell\left(\sum_{n=1}^{\ell}\left(-b_{n}^{(k)}+c_{n}^{(k)}\right)+N\ell e_{k}+\left(N-1\right)\ell\left(g-1\right)\right)\right]\times\nonumber\\
 & & PE\left[\left(-\sum_{n=1}^{\ell}b_{n}^{(k)}+\ell c_{k}^{(k)}+N\ell e_{k}+\left(N-1\right)\ell\left(g-1\right)\right)\right]\times\nonumber\\
 & & PE\left[\sum_{n=1}^{\ell}\sum_{j=\ell+1}^{k-1}\left(-b_{n}^{(k)}+c_{j}^{(k)}+Ne_{k}+\left(N-1\right)\left(g-1\right)\right)\widetilde{\gamma}_{k-\ell}^{-N}\widetilde{\gamma}_{j-\ell}^{N}\right]\times\nonumber\\
 & & PE\left[\sum_{i=\ell+1}^{k}\sum_{m=1}^{\ell}\left(-b_{i}^{(k)}+c_{m}^{(k)}+Ne_{k}+\left(N-1\right)\left(g-1\right)\right)\widetilde{\beta}_{i-\ell}^{-N}\widetilde{\gamma}_{k-\ell}^{N}T^{Nk/\left(k-\ell\right)}\right]\times\nonumber\\
 & & PE\left[\sum_{i=\ell+1}^{k}\sum_{j=\ell+1}^{k}\left(-b_{i}^{(k)}+c_{j}^{(k)}+Ne_{k}+\left(N-1\right)\left(g-1\right)\right)\widetilde{\beta}_{i-\ell}^{-N}\widetilde{\gamma}_{j-\ell}^{N}T^{Nk/\left(k-\ell\right)}\right]\times\nonumber\\
 & & PE\left[\sum_{n=1}^{N-1}\left(kne_{k}+\left(kn-1\right)\left(g-1\right)\right)T^{kn}\right]\,.
\ee
As before we need to redefine the t- and R-charges, in this case such that $T\rightarrow T^{(k-\ell)/k}$. The result is
\be
\mathcal{I}_{g,\left(b_{i},c_{j},e\right)}^{N,k(flow)} & = & PE\left[\sum_{j=1}^{k-\ell-1}\left(-\sum_{n=1}^{\ell}b_{n}^{(k)}+\ell\left(c_{j+\ell}^{(k)}+Ne_{k}+\left(N-1\right)\left(g-1\right)\right)\right)\widetilde{\gamma}_{k-\ell}^{-N}\widetilde{\gamma}_{j}^{N}\right]\times\nonumber\\
 & & PE\left[\sum_{i=1}^{k-\ell}\left(\sum_{m=1}^{\ell}c_{m}^{(k)}+\ell\left(-b_{i+\ell}^{(k)}+Ne_{k}+\left(N-1\right)\left(g-1\right)\right)\right)\widetilde{\beta}_{i}^{-N}\widetilde{\gamma}_{k-\ell}^{N}T^{N}\right]\times\nonumber\\
 & & PE\left[\sum_{i=1}^{k-\ell}\sum_{j=1}^{k-\ell}\left(-b_{i+\ell}^{(k)}+c_{j+\ell}^{(k)}+Ne_{k}+\left(N-1\right)\left(g-1\right)\right)\widetilde{\beta}_{i}^{-N}\widetilde{\gamma}_{j}^{N}T^{N}\right]\times\nonumber\\
 & & PE\left[\sum_{n=1}^{N-1}\left(kne_{k}+\left(kn-1\right)\left(g-1\right)\right)T^{\left(k-\ell\right)n}\right],
\ee
where as before we stripped the divergence. We can now compare the above result to the Coulomb limit formula of class $\mathcal{S}_{k-\ell}$
\be
\mathcal{I}_{g,\left(b_{i},c_{j},e\right)}^{N,k-\ell} & = & PE\left[\sum_{i,j=1}^{k-\ell}\left(-b_{i}^{\left(k-\ell\right)}+c_{j}^{\left(k-\ell\right)}+Ne_{k-\ell}+\left(N-1\right)\left(g-1\right)\right)\beta_{i}^{-N}\gamma_{j}^{N}T^{N}\right]\times\nonumber\\
 & & PE\left[\sum_{n=1}^{N-1}\left(\left(k-\ell\right)ne_{k-\ell}+\left(\left(k-\ell\right)n-1\right)\left(g-1\right)\right)T^{\left(k-\ell\right)n}\right]\,.
\ee

We can match all the coefficients of $T^{(k-\ell)n}$ by setting the t-flux of the $\mathcal{S}_{k-\ell}$ theory to be
\be
\label{E:tFluxGen}
e_{k-\ell}=\frac{k}{k-\ell}e_{k}+\frac{\ell}{k-\ell}\left(g-1\right)\,.
\ee
Comparing the coefficients of $\beta_{i}^{-N}\gamma_{j}^{N}T^{N}$ for $i=1,...,k-\ell$ and $j=1,...,k-\ell-1$ we find the $\beta$- and $\gamma$-fluxes of the $\mathcal{S}_{k-\ell}$ theory need to be set to
\be
\label{E:bcFluxGen}
\label{E:beta-gamma fluxes}
b_{i}^{\left(k-\ell\right)} & = & b_{i+\ell}^{(k)}+\frac{1}{k-\ell}\sum_{n=1}^{\ell}b_{n}^{(k)}\,, \qquad i=1,...,k-\ell\nonumber\\
c_{j}^{\left(k-\ell\right)} & = & c_{j+\ell}^{(k)}+\frac{1}{k-\ell}\left(\sum_{i=1}^{\ell}b_{i}^{(k)}-N\ell e_{k}-N\ell\left(g-1\right)\right)\,, \qquad j=1,...,k-\ell-1\,.
\ee

In this case there will be additional free fields that decouple in the flow, associated with the unmatched coefficients between the two Coulomb limits. The first is the residual coefficient of $\widetilde{\beta}_i^{-N}\widetilde{\gamma}_{k-\ell}^{N}T^N$ where $i=1,...,k-\ell$. Those fields have $-N$ $\widetilde{\beta}_i$ charge, $N$ $\widetilde{\gamma}_{k-\ell}$ charge, $2N$ R-charge, and $-N$ t-charge. The number of such free chiral multiplets
\be
n_{FC,i}\left(\widetilde{\beta}_i^{-N}\widetilde{\gamma}_{k-\ell}^{N}T^N\right) = \sum_{n=1}^{\ell}b_{n}^{(k)}-\ell b_{i+1}^{(k)}+\ell\left(g-1\right)\,.
\ee
The second free field emerging from the coefficient of $\widetilde{\gamma}_{j}^{N}\widetilde{\gamma}_{k-\ell}^{-N}$, where $j=1,...,k-\ell-1$, with $2N$ $\widetilde{\gamma}_{j}$-charge and $N$ charge units for all the other $\widetilde{\gamma}$-s (This is due to the relation $\widetilde{\gamma}_{k-\ell}=\prod_{j=1}^{k-\ell-1}\widetilde{\gamma}_j^{-1}$). These free chiral fields are numerated by
\be
n_{FC,j}\left(\widetilde{\gamma}_{j}^{N}\widetilde{\gamma}_{k-\ell}^{-N}\right) = -\sum_{n=1}^{\ell}b_{n}^{(k)}+\ell\left(c_{j+\ell}^{(k)}+Ne_{k}+\left(N-1\right)\left(g-1\right)\right)\,.
\ee

Many of these results were chosen in a manner that match the direct results from Lagrangian theories flows, and the anomaly polynomial flows we will present in the following sections. Meaning, these results cannot be determined fully by the Coulomb limit comparisons. One simple check we can make is by taking $\ell=k-1$; this restores the former case of breaking the $\beta$ and $\gamma$ symmetries completely. In addition, as before one may use other vevs to eliminate partially the internal symmetries and reach class $\mathcal{S}_{k-\ell}$, but all of these can be trivially mapped to the case studied above.

\subsection{Flows from class $\mathcal{S}_k$ to class $\mathcal{S}_{k-\ell}$ with extra punctures}
\label{SS:Coulomb Flow W Punct}

Finally, we will employ the Coulomb index to find the mapping of fluxes and new minimal punctures in the case we give vevs that break some of the $U(1)_{\beta_i}\times U(1)_{\gamma_j}$ symmetries to diagonal $U(1)$ symmetries. The initial Coulomb limit will be of puncture-less Riemann surfaces, while class $\mathcal{S}_{k-\ell}$ Coulomb limit will be of Riemann surfaces with negative minimal punctures. For the conventions of punctures in class $\mathcal{S}_k$ see Appendix \ref{A:PuncConventions}.

The vevs are chosen in a rather general form breaking $\ell$ pairs of $U(1)_{\beta_n}\times U(1)_{\gamma_n}$ to $\ell$ $U(1)$ factors. The vevs are $\left(\beta_{n}^{-1}\gamma_{n}\left(\frac{pq}{t}\right)\right)^{N}=1$ with $n=1,...,\ell$, translating to the assignments
\be
\label{E:General vevs min}
\beta_{n}=\left(\frac{pq}{t}\right)^{1/2}\epsilon_{n}^{-1}\prod_{r=1}^{\ell}\epsilon_{r}^{1/k}\,,&\;\beta_{i}=\widetilde{\beta}_{i-\ell}\left(\frac{pq}{t}\right)^{-\ell/2\left(k-\ell\right)}\prod_{r=1}^{\ell}\epsilon_{r}^{1/k}\qquad &n=1,...,\ell\nonumber\\
\gamma_{n}=\left(\frac{pq}{t}\right)^{-1/2}\epsilon_{n}^{-1}\prod_{r=1}^{\ell}\epsilon_{r}^{1/k}\,,&\;\gamma_{i}=\widetilde{\gamma}_{i-\ell}\left(\frac{pq}{t}\right)^{\ell/2\left(k-\ell\right)}\prod_{r=1}^{\ell}\epsilon_{r}^{1/k}\qquad &i=\ell+1,...,k\,,
\ee
where $\epsilon_n$ is the fugacity matching the residual $U(1)_{\epsilon_n}$ from the breaking of $U(1)_{\beta_n}\times U(1)_{\gamma_n}$ by the vevs. Also, we have mixed the $U(1)$ factors in a specific manner to make the minimal punctures more apparent.

Using the above assignments in the Coulomb limit formula \eqref{E:Coulomb}, reproduced here
\be
\mathcal{I}_{g,\left(b_{i},c_{j},e\right)}^{N,k} & = & PE\left[\sum_{i,j=1}^{k}\left(-b_{i}+c_{j}+Ne_{k}+\left(N-1\right)\left(g-1\right)\right)\beta_{i}^{-N}\gamma_{j}^{N}T^{N}\right]\times\nonumber\\
 & & PE\left[\sum_{n=1}^{N-1}\left(kne_{k}+\left(kn-1\right)\left(g-1\right)\right)T^{kn}\right]\,,
\ee
it transforms as
\be
\mathcal{I}_{g,\left(b_{i},c_{j},e\right)}^{N,k(flow)} & = & PE\left[\sum_{n=1}^{\ell}\left(-b_{n}^{\left(k\right)}+c_{n}^{\left(k\right)}\right)+N\ell e_{k}+\left(N-1\right)\ell\left(g-1\right)\right]\times\nonumber\\
 & & PE\left[\sum_{n\ne m}^{\ell}\left(-b_{n}^{\left(k\right)}+c_{m}^{\left(k\right)}+Ne_{k}+\left(N-1\right)\left(g-1\right)\right)\epsilon_{n}^{N}\epsilon_{m}^{-N}\right]\times\nonumber\\
 & & PE\left[\sum_{n=1}^{\ell}\sum_{j=\ell+1}^{k}\left(-b_{n}^{\left(k\right)}+c_{j}^{\left(k\right)}+Ne_{k}+\left(N-1\right)\left(g-1\right)\right)\widetilde{\gamma}_{j-\ell}^{N}\epsilon_{n}^{N}T^{Nk/2\left(k-\ell\right)}\right]\times\nonumber\\
 & & PE\left[\sum_{i=\ell+1}^{k}\sum_{m=1}^{\ell}\left(-b_{i}^{\left(k\right)}+c_{m}^{\left(k\right)}+Ne_{k}+\left(N-1\right)\left(g-1\right)\right)\widetilde{\beta}_{i-\ell}^{-N}\epsilon_{m}^{-N}T^{Nk/2\left(k-\ell\right)}\right]\times\nonumber\\
 & & PE\left[\sum_{i,j=\ell+1}^{k}\left(-b_{i}^{\left(k\right)}+c_{j}^{\left(k\right)}+Ne_{k}+\left(N-1\right)\left(g-1\right)\right)\widetilde{\beta}_{i-\ell}^{-N}\widetilde{\gamma}_{j-\ell}^{N}T^{Nk/\left(k-\ell\right)}\right]\times\nonumber\\
 & & PE\left[\sum_{n=1}^{N-1}\left(kne_{k}+\left(kn-1\right)\left(g-1\right)\right)T^{kn}\right]\, .
\ee
Redefining the t- and R-charges s.t. $T\rightarrow T^{\left(k-\ell\right)/k}$ results in
\be
\mathcal{I}_{g,\left(b_{i},c_{j},e\right)}^{N,k(flow)} & = & PE\left[\sum_{n\ne m}^{\ell}\left(-b_{n}^{\left(k\right)}+c_{m}^{\left(k\right)}+Ne_{k}+\left(N-1\right)\left(g-1\right)\right)\epsilon_{n}^{N}\epsilon_{m}^{-N}\right]\times\nonumber\\
 & & PE\left[\sum_{n=1}^{\ell}\sum_{j=1}^{k-\ell}\left(-b_{n}^{\left(k\right)}+c_{j+\ell}^{\left(k\right)}+Ne_{k}+\left(N-1\right)\left(g-1\right)\right)\widetilde{\gamma}_{j}^{N}\epsilon_{n}^{N}T^{N/2}\right]\times\nonumber\\
 & & PE\left[\sum_{i=1}^{k-\ell}\sum_{n=1}^{\ell}\left(-b_{i+\ell}^{\left(k\right)}+c_{n}^{\left(k\right)}+Ne_{k}+\left(N-1\right)\left(g-1\right)\right)\widetilde{\beta}_{i}^{-N}\epsilon_{n}^{-N}T^{N/2}\right]\times\nonumber\\
 & & PE\left[\sum_{i,j=1}^{k-\ell}\left(-b_{i+\ell}^{\left(k\right)}+c_{j+\ell}^{\left(k\right)}+Ne_{k}+\left(N-1\right)\left(g-1\right)\right)\widetilde{\beta}_{i}^{-N}\widetilde{\gamma}_{j}^{N}T^{N}\right]\times\nonumber\\
 & & PE\left[\sum_{n=1}^{N-1}\left(kne_{k}+\left(kn-1\right)\left(g-1\right)\right)T^{\left(k-\ell\right)n}\right]\,,
\ee
where we removed the divergence.

This result needs to be compared with the Coulomb limit formula of class $\mathcal{S}_{k-\ell}$ of a Riemann surface with $m_n^-$ negative minimal punctures charged under $U(1)_{\epsilon_n}$ where $n=1,...,\ell$
\be
\mathcal{I}_{g,m_{n}^{-},\left(b_{i},c_{j},e\right)}^{N,k-\ell} & = & PE\left[\sum_{i,j=1}^{k-\ell}\left(-b_{i}^{\left(k-\ell\right)}+c_{j}^{\left(k-\ell\right)}+N\left(e_{k-\ell}+\frac{m_{tot}^{-}}{2\left(k-\ell\right)}\right)\right)\beta_{i}^{-N}\gamma_{j}^{N}T^{N}\right]\times\nonumber\\
 & & PE\left[\sum_{i,j=1}^{k-\ell}\left(\left(N-1\right)\left(g-1\right)-\frac{m_{tot}^{-}}{k-\ell}\right)\beta_{i}^{-N}\gamma_{j}^{N}T^{N}\right]\times\nonumber\\
 & & PE\left[\sum_{n=1}^{N-1}\left(\left(k-\ell\right)n\left(e_{k-\ell}+\frac{m_{tot}^{-}}{2\left(k-\ell\right)}\right)+\left(\left(k-\ell\right)n-1\right)\left(g-1\right)\right)T^{\left(k-\ell\right)n}\right]\nonumber\\
 & & PE\left[\sum_{i=1}^{k-\ell}\sum_{n=1}^{\ell}\left(m_{n}^{-}\right)\left(\beta_{i}^{-N}\epsilon_{n}^{-N}+\gamma_{i}^{N}\epsilon_{n}^{N}\right)T^{N/2}\right]\,,
\ee
where again $m_n^-$ is the number of negative minimal punctures associated to the symmetry $U(1)_{\epsilon_n}$, and $m_{tot}^-$ is the total number of negative minimal punctures. Matching the coefficients of $T^{\left(k-\ell\right)n}$ we find the t-flux of the resulting $\mathcal{S}_{k-\ell}$ theory is
\be
\label{E:tFluxGen min}
e_{k-\ell}=\frac{k}{k-\ell}e_{k}+\frac{\ell}{k-\ell}\left(g-1\right)-\frac{m_{tot}^{-}}{2\left(k-\ell\right)}\,.
\ee      
This result reduces to the one we found with the initial set of vevs we used in equation \eqref{E:tFluxGen} when we close all the minimal punctures.\footnote{Recall that each closing of a negative minimal puncture in class $\mathcal{S}_k$ shifts the $t$-flux by $+\frac{1}{2k}$.}

Next, we can compare the coefficients of $\beta_{i}^{-N}\gamma_{j}^{N}T^{N}$ for $i,j=1,...,k-\ell$. The resulting $\beta$- and $\gamma$-fluxes of the IR theory
\be
\label{E:beta-gamma fluxes min}
b_{i}^{\left(k-\ell\right)} & = & b_{i+\ell}^{(k)}+\frac{1}{k-\ell}\sum_{n=1}^{\ell}b_{n}^{(k)}\qquad i=1,...,k-\ell\nonumber\\
c_{j}^{\left(k-\ell\right)} & = & c_{j+\ell}^{(k)}+\frac{1}{k-\ell}\sum_{n=1}^{\ell}c_{n}^{(k)}\qquad j=1,...,k-\ell\,,
\ee
where the symmetry between the $\beta$-s and $\gamma$-s was required from the symmetry of the given vevs. These formulas show that the residual flux from the $\ell$ broken $U(1)_{\beta_n}$-s and $U(1)_{\gamma_m}$-s is divided equally between the $(k-\ell)$ remaining $U(1)_{\beta_i}$-s and $U(1)_{\gamma_j}$-s in a manner that preserves the required relations $\prod_i \beta_i =1$ and $\prod_j \gamma_i =1$.

The results we found for the $\beta$- and $\gamma$-fluxes need to be consistent with the results we found with our initial set of vevs when we close all the minimal punctures with appropriate vevs. From this consistency requirement we can find the number of additional minimal punctures, since it should be equal to the difference between the $\gamma_k$ flux found in the above formula \eqref{E:beta-gamma fluxes min} and the one found with the initial set of vevs in \eqref{E:beta-gamma fluxes} up to a normalization.\footnote{Remember that closing of a negative minimal puncture of class $\mathcal{S}_k$ with vev containing $\gamma_k$ shifts the $\gamma_k$ flux by $+\frac{k-1}{k}$} This results in the number of new (negative) minimal punctures being
\be
m_{tot}^{-}=\sum_{n=1}^{\ell}c_{n}^{(k)}-\sum_{n=1}^{\ell}b_{n}^{(k)}+N\ell e_{k}+N\ell\left(g-1\right)\,. \label{E:PuncNumber}
\ee
One can consider the vevs we gave as a series of vevs each decreasing $k$ by $1$. This allows us to calculate the number of minimal punctures of each kind from the above formula alone. The resulting number of additional minimal punctures relating to each $U(1)_{\epsilon_n}$ is
\be
\label{E:minPunc}
m_{n}^{-}=c_{n}^{(k)}-b_{n}^{(k)}+Ne_{k}+N\left(g-1\right)\,.
\ee

To match the two Coulomb branch formulas completely one needs to remove some free chiral fields that decouple during the flow as we had before. We find decoupled free chirals from the residual coefficients of $\widetilde{\beta}_{i}^{-N}\epsilon_{n}^{-N}T^{N/2}$ and $\widetilde{\gamma}_{j}^{N}\epsilon_{n}^{N}T^{N/2}$ with $i,j=1,...,k-\ell$ and $n=1,...,\ell$. These fields have $-N$ $\widetilde{\beta}_i$ and $N$ $\widetilde{\gamma}_j$ charge, respectively, $-N$ and $+N$ $\epsilon_n$ charge, respectively. In addition they have $N$ R-charge, and $-N/2$ t-charge. numerated by
\be
\label{E:FreeChirals min}
n_{FC}\left(\widetilde{\beta}_{i}^{-N}\epsilon_{n}^{-N}T^{N/2}\right) & = & -b_{i+\ell}^{\left(k\right)}+b_{n}^{(k)}-\left(g-1\right)\nonumber\\
n_{FC}\left(\widetilde{\gamma}_{j}^{N}\epsilon_{n}^{N}T^{N/2}\right) & = & c_{j+\ell}^{\left(k\right)}-c_{n}^{(k)}-\left(g-1\right)\,.
\ee
These are the same free chiral fields shown in equation \eqref{E:FC6d} that were expected from the $6d$ flow. Here we see that in the $4d$ flow the number of these free chirals is effected by the flux used in the compactification.

Additional decoupled free chirals come from the coefficient of $\epsilon_{n}^{N}\epsilon_{m}^{-N}$ with $n\ne m=1,...,\ell$, and charges $+N$ and $-N$ for $\epsilon_n$ and $\epsilon_m$. The number of such free chirals
\be
n_{FC}\left(\epsilon_{n}^{N}\epsilon_{m}^{-N}\right) = -b_{n}^{\left(k\right)}+c_{m}^{\left(k\right)}+Ne_{k}+\left(N-1\right)\left(g-1\right)\,.
\ee
These free chirals will not arise in the case we give a vev to only one operator. In fact they will not appear at all if one sequentially give vev to one operator at a time and decouples its free fields before going to the next.

\section{Full index computation}
\label{S:IndexComp}
In this section we will focus on explicit examples of theories in class $\mathcal{S}_k$ that have a known Lagrangian description, allowing us to write for them a closed form superconformal index. We will initiate the flow on the level of the index and try to identify the resulting index in class $\mathcal{S}_{k'}$. We give the relevant index definitions in Appendix \ref{A:indexdefinitions}.

\subsection{Flows from $\mathcal{S}_2$ to $\mathcal{S}_1$ with $N=2$}

In the case of classes $\mathcal{S}_2$ and $\mathcal{S}_1$ with $N=2$ we know the Lagrangians of all theories. This enables to do explicit index calculations to further support the results from the Coulomb limit formula. In addition, these calculation will enable to further establish the claim on the free chiral multiplets that decouple during the flow. We will consider flows from theories described by a genus $g=0,1,2$ Riemann surface and some flux that will contain all poles matching the vacuum expectation values (vevs) that appear above \eqref{E:CoulombFlow}. We will use these index flows to verify the results of both the case with additional minimal punctures and the one without. In all cases we will start from giving a vev to only one baryonic operator given by
\be
\label{E:vev N=k=2}
\left(\beta^{-1}\gamma\left(\frac{pq}{t}\right)\right)^{2}=1\, .
\ee 
This translates to the assignments $\beta=\left(\frac{pq}{t}\right)^{1/2}\epsilon^{-1/2}$ and $\gamma=\left(\frac{pq}{t}\right)^{-1/2}\epsilon^{-1/2}$, leaving us with additional minimal punctures of fugacity $\epsilon$. The resulting indices from this flow will allow us to verify the results of Subsection \ref{SS:Coulomb Flow W Punct}. The second baryonic operator vev used in Subsection \ref{SS:Coulomb Flow WO Punct N=k=2} is
\be
\left(\beta^{-1}\gamma^{-1}\left(\frac{pq}{t}\right)\right)^{2}=1\, ,
\ee
and it breaks all the $\beta$ and $\gamma$ internal symmetries completely. Plugging the assignments of the former vev together with the identifications of the new $t$- and $R$- charges involved in the flow, we find it translates to
\be
\left(\epsilon\sqrt{\frac{pq}{t}}\right)^{2}=1\, .
\ee
This is exactly the vev that one gives to a baryonic operator charged under a $U(1)$ symmetry associated to a negative minimal puncture with fugacity $\epsilon$ in order to close the puncture. We will use this second flow to verify the results of Subsection \ref{SS:Coulomb Flow WO Punct N=k=2}.

\subsubsection*{Spheres ($g=0$)}
To generate the index of theories described by a sphere one can simply take a free trinion\footnote{A free trinion is a theory described by compactification on a sphere with two maximal punctures and one minimal puncture.} with t-flux tubes glued to it (see appandix of \cite{Bah:2017gph}), and close all the punctures.

The first example is of an $N=k=2$ sphere with t-flux $e=1$ and additional $\beta$-flux of $b=-1$ and $\gamma$-flux of $c=-1$ 
\be
\mathcal{I}_{g=0,m^-=0,\left(e=1,b=-1,c=-1\right)}^{N=2,k=2} & = & \Gamma_{e}\left(pq\beta^{-4}\right)\Gamma_{e}\left(pq\gamma^{-4}\right)\Gamma_{e}\left(\left(\beta^{-1}\gamma\right)^{\pm2}\right)\Gamma_{e}\left(\left(\frac{pq}{t}\right)^{2}\left(\beta^{-1}\gamma\right)^{\pm2}\right)\times\nonumber\\
 & & \Gamma_{e}\left(\left(\left(\frac{pq}{t}\right)^{2}\beta^{-2}\gamma^{-2}\right)^{\pm1}\right)\Gamma_{e}\left(\left(\frac{pq}{t}\right)^{2}\beta^{-2}\gamma^{-2}\right)^{2}\, .
\ee
Both required poles appear, and we can initiate the flow by giving the first vev. In this case it's trivial because the above index describes a WZ model. Thus, simply some of the fields become massive and decouple in the IR. After the required t- and R-charge redefinition $\left(\frac{pq}{t}\right)\rightarrow\left(\frac{pq}{t}\right)^{1/2}$ we find
\be
\mathcal{I}_{g=0,m^-=0,\left(e=1,b=-1,c=-1\right)}^{N=2,k=2,flow\ 1} & = & \Gamma_{e}\left(\left(\frac{pq}{t}\right)^{2}\right)\Gamma_{e}\left(\left(\frac{pq}{t}\right)^{\pm1}\right)\Gamma_{e}\left(\frac{pq}{t}\epsilon^{2}\right)^{3}\Gamma_{e}\left(\frac{pq}{t}\epsilon^{-2}\right)^{-1}\nonumber\\
 & = & \Gamma_{e}\left(\frac{pq}{t}\epsilon^{2}\right)^{3}\Gamma_{e}\left(\frac{pq}{t}\epsilon^{-2}\right)^{-1}\mathcal{I}_{g=0,m^-=0,\left(e=1\right)}^{N=2,k=1}\, ,
\ee
where $\mathcal{I}_{g=0,m^-=0,\left(e=1\right)}^{N=2,k=1}$ is the matching $\mathcal{S}_1$ index, with the expected number of minimal punctures $m^-=-1 - (-1) + 2\cdot 1 + 2(0-1)=0$  and t-flux $e=2\cdot 1+(2-1)(0-1)-0=1$ predicted by \eqref{E:minPunc} and \eqref{E:tFluxGen min}. In addition we find the expected free chirals contribution is as predicted in \eqref{E:FreeChirals min}, $n_{FC}\left(\epsilon^{-2}T\right)=-1 - 1 - (0-1)=-1$ and $n_{FC}\left(\epsilon^{2}T\right)=-(-1) - (-1) - (0-1)=3$.

Next, we give a vev to the second operator setting $\epsilon=\left(\frac{pq}{t}\right)^{-1/2}$. This only effects the free chirals since we got no extra minimal punctures, and we get
\be
\mathcal{I}_{g=0,m^-=0,\left(e=1,b=-1,c=-1\right)}^{N=2,k=2,flow\ 2} = \Gamma_{e}\left(\left(\frac{pq}{t}\right)^{\pm1}\right)
 = \Gamma_{e}\left(\left(\frac{pq}{t}\right)^{2}\right)^{-1}\mathcal{I}_{\left(e=1\right)}^{N=2,k=1}\, ,
\ee
where we find the expected free chirals contribution predicted in \eqref{E:freeChirals} $n_{FC}=-2\cdot 1 - (2-1)(0-1)=-1$. The rest of the parameters of the result match in the same way as before.

The second example is of a sphere with t-flux $e=2$ and vanishing $\beta$- and $\gamma$-flux 
\be
\mathcal{I}_{\left(e=2,b=0,c=0\right)}^{N=2,k=2} & = & \Gamma_{e}\left(\left(\frac{pq}{t}\right)^{2}\right)^{2}\Gamma_{e}\left(\left(\frac{pq}{t}\right)^{2}\left(\beta\gamma\right)^{\pm2}\right)\times\nonumber\\
 & & \Gamma_{e}\left(\left(\frac{pq}{t}\right)^{2}\left(\beta\gamma^{-1}\right)^{\pm2}\right)^{2}\mathcal{I}^{t-tube}\left(\left\{ \frac{pq}{t},\frac{\gamma}{\beta}\right\} ,\left\{ \frac{\gamma}{\beta},\frac{pq}{t}\right\} \right)\, ,
\ee
where
\be
\mathcal{I}^{t-tube}\left(\mathbf{v},\mathbf{c}\right) & = & \Gamma_{e}\left(t\left(\gamma\beta^{-1}v_{2}\right)^{\pm1}v_{1}^{\pm1}\right)\Gamma_{e}\left(pq\gamma^{-2}\beta^{-2}\right)\Gamma_{e}\left(\left(\frac{pq}{t}\right)^{2}\right)\Gamma_{e}\left(\left(\frac{pq}{t}\right)^{2}\left(\gamma\beta^{-1}\right)^{\pm2}\right)\times\nonumber\\
 & & \Gamma_{e}\left(\left(\frac{pq}{t}\right)^{2}\left(\gamma\beta\right)^{2}\right)\kappa\oint\frac{dz}{4\pi iz}\frac{\Gamma_{e}\left(\frac{pq}{t\gamma\beta}\left(\beta\gamma^{-1}v_{2}^{-1}\right)^{\pm1}z^{\pm1}\right)}{\Gamma_{e}\left(z^{\pm2}\right)}\Gamma_{e}\left(\gamma\beta z^{\pm1}v_{1}^{\pm1}\right)\times\nonumber\\
 & & \mathcal{I}^{orbifold}_{\left\{ \beta\gamma^{-1},\frac{pq}{t}\right\}, \mathbf{c},\sqrt{zv_{2}},\sqrt{v_{2}/z}}\, .
\ee
The orbifold theory is the theory of two free trinions $\Phi$-glued together (See Appendix \ref{A:PuncConventions}), with index
\be
\mathcal{I}_{\mathbf{z},\mathbf{c},a,b}^{orbifold} & = & \kappa^{2}\oint\frac{dw_{1}}{4\pi iw_{1}}\oint\frac{dw_{2}}{4\pi iw_{2}}\frac{\Gamma_{e}\left(\frac{pq}{t}\left(\beta\gamma\right)^{\pm1}w_{1}^{\pm1}w_{2}^{\pm1}\right)}{\Gamma_{e}\left(w_{1}^{\pm2}\right)\Gamma_{e}\left(w_{2}^{\pm2}\right)}\nonumber\\
 & & \Gamma_{e}\left(t^{\frac{1}{2}}\beta a^{-1}w_{1}^{\pm1}z_{1}^{\pm1}\right)\Gamma_{e}\left(t^{\frac{1}{2}}\gamma^{-1}aw_{1}^{\pm1}z_{2}^{\pm1}\right)\Gamma_{e}\left(t^{\frac{1}{2}}\gamma aw_{2}^{\pm1}z_{1}^{\pm1}\right)\times\nonumber\\
 & & \Gamma_{e}\left(t^{\frac{1}{2}}\beta^{-1}a^{-1}w_{2}^{\pm1}z_{2}^{\pm1}\right)\Gamma_{e}\left(t^{\frac{1}{2}}\gamma bw_{1}^{\pm1}c_{1}^{\pm1}\right)\Gamma_{e}\left(t^{\frac{1}{2}}\beta^{-1}b^{-1}w_{1}^{\pm1}c_{2}^{\pm1}\right)\times\nonumber\\
  & & \Gamma_{e}\left(t^{\frac{1}{2}}\beta b^{-1}w_{2}^{\pm1}c_{1}^{\pm1}\right)\Gamma_{e}\left(t^{\frac{1}{2}}\gamma^{-1}bw_{2}^{\pm1}c_{2}^{\pm1}\right)\, .
\ee
The vev we give in \eqref{E:vev N=k=2} generates an RG-flow going from the UV theory in high energies to the IR theory in low energies. Going to lower energies than the energy scale associated with the vev, the vacuum becomes none invariant under the gauge symmetry, effectively breaking it. This procedure is known as the Higgs mechanism, and we will refer to gauge symmetries broken in such a manner as "Higgsed". In the above example one of the three $SU(2)$ gauge symmetries is Higgsed during the flow, and one of the gauge symmetries is left with three flavors; thus, described in the IR by quadratic gauge invariant composites. additionally, several fields become massive and decouple, resulting in
\be
\mathcal{I}_{g=0,m^-=0\left(e=2,b=0,c=0\right)}^{N=2,k=2,flow\ 1} & = & \Gamma_{e}\left(\frac{pq}{t}\right)^{2}\Gamma_{e}\left(\frac{pq}{t}\epsilon^{\pm2}\right)^{3}\Gamma_{e}\left(\left(\frac{pq}{t}\right)^{2}\right)^{2}\nonumber\\
 & & \kappa\oint\frac{dw_{2}}{4\pi iw_{2}}\frac{\Gamma_{e}\left(\frac{pq}{t}\right)^{2}\Gamma_{e}\left(\frac{pq}{t}w_{2}^{\pm2}\right)}{\Gamma_{e}\left(w_{2}^{\pm2}\right)}\Gamma_{e}\left(\frac{t}{\sqrt{pq}}\epsilon^{\pm1}w_{2}^{\pm1}\right)^{2}\nonumber\\
 & = & \Gamma_{e}\left(\frac{pq}{t}\epsilon^{\pm2}\right)\mathcal{I}_{g=0,m^{-}=2,\left(e=2\right)}^{N=2,k=1}
\ee
where again we find the expected  number of additional minimal punctures $m^-=0 - 0 + 2\cdot 2 + 2(0-1)=2$, $t$-flux $e=2\cdot 2+(2-1)(0-1)-2/2=2$ and free chirals $n_{FC}\left(\epsilon^{-2}T\right)= - (0-1)=1$ and $n_{FC}\left(\epsilon^{2}T\right)= - (0-1)=1$.

Now, as before we give a vev to the second operator setting $\epsilon=\left(\frac{pq}{t}\right)^{-1/2}$. This closes all the minimal punctures, and we remain with
\be
\mathcal{I}_{g=0,m^-=0,\left(e=2,b=0,c=0\right)}^{N=2,k=2,flow\ 2} & = & \Gamma_{e}\left(\frac{pq}{t}\right)^{4}\Gamma_{e}\left(\left(\frac{pq}{t}\right)^{2}\right)^{5}\times\nonumber\\
 & & \kappa\oint\frac{dw}{4\pi iw}\frac{\Gamma_{e}\left(\frac{pq}{t}w^{\pm2}\right)}{\Gamma_{e}\left(w^{\pm2}\right)}\Gamma_{e}\left(\frac{t^{3/2}}{pq}w^{\pm1}\right)^{2}\Gamma_{e}\left(t^{1/2}w^{\pm1}\right)^{2}\nonumber\\
 & = & \Gamma_{e}\left(\left(\frac{pq}{t}\right)^{2}\right)\mathcal{I}_{\left(e=3\right)}^{N=2,k=1}\,,
\ee
where the result parameters are as expected $e=2\cdot 2+(2-1)(0-1)=3$ and $n_{FC}=-2\cdot 0 - (2-1)(0-1)=1$.

\subsubsection*{Tori ($g=1$)}
The first torus theory we consider has t-flux $e=0$, with $\beta$- and $\gamma$-flux $b=-1$ and $c=0$, chosen such that the index contains the required poles 
\be
\mathcal{I}_{g=1,m^-=0,\left(e=0,b=-1,c=0\right)}^{N=2,k=2} & = & \kappa^{2}\oint\frac{du_{1}}{4\pi iu_{1}}\oint\frac{du_{2}}{4\pi iu_{2}}\frac{\Gamma_{e}\left(\frac{pq}{t}\left(\beta\gamma^{-1}\right)^{\pm1}u_{1}^{\pm1}u_{2}^{\pm1}\right)}{\Gamma_{e}\left(u_{1}^{\pm2}\right)\Gamma_{e}\left(u_{2}^{\pm2}\right)}\Gamma_{e}\left(pq\beta^{-4}\right)^{2}\times\nonumber\\
 & & \Gamma_{e}\left(u_{1}^{\pm1}u_{1}^{\pm1}\beta^{2}\right)\Gamma_{e}\left(tu_{2}^{\pm1}u_{1}^{\pm1}\gamma\beta^{-1}\right)\Gamma_{e}\left(tu_{1}^{\pm1}u_{2}^{\pm1}\gamma\beta^{-1}\right)\times\nonumber\\
 & & \Gamma_{e}\left(u_{2}^{\pm1}u_{2}^{\pm1}\beta^{2}\right)\frac{\Gamma_{e}\left(tu_{2}^{\pm1}u_{1}^{\pm1}\left(\gamma\beta\right)^{-1}\right)}{\Gamma_{e}\left(tu_{2}^{\pm1}u_{1}^{\pm1}\left(\gamma\beta\right)\right)}\,.
\ee
The flow leads to a Higgsing of one of the $SU(2)$ gauge symmetries, the remaining index is thus
\be
\mathcal{I}_{g=1,m^-=0,\left(e=0,b=-1,c=0\right)}^{N=2,k=2,flow\ 1} & = & \Gamma_{e}\left(\frac{pq}{t}\epsilon^{-2}\right)^{-2}\kappa\oint\frac{du}{4\pi iu}\frac{\Gamma_{e}\left(tu^{\pm1}u^{\pm1}\right)}{\Gamma_{e}\left(u^{\pm2}\right)}\Gamma_{e}\left(\sqrt{\frac{pq}{t}}u^{\pm1}u^{\pm1}\epsilon^{\pm1}\right)\nonumber\\
 & = & \Gamma_{e}\left(\frac{pq}{t}\epsilon^{-2}\right)^{-2}\mathcal{I}_{g=1,m^{-}=1,\left(e=-1/2\right)}^{N=2,k=1}
\ee
with expected additional minimal punctures, flux and free chirals.

Giving the second vev, we find that the $SU(2)$ gauge is removed since all the remaining multiplets that transform under it become massive. The resulting index of the IR theory is
\be
\mathcal{I}_{g=1,m^-=0,\left(e=0,b=-1,c=0\right)}^{N=2,k=2,flow\ 2} & = & \Gamma_{e}\left(\left(\frac{pq}{t}\right)^{2}\right)^{-2}\frac{\kappa}{2}=\Gamma_{e}\left(\left(\frac{pq}{t}\right)^{2}\right)^{-2}\mathcal{I}_{g=1,m^-=0,\left(e=0\right)}^{N=2,k=1}\,,
\ee
with expected free chirals and flux. Notice that the remaining theory in class $\mathcal{S}$ is described by a torus compactification with no flux, so it actually has $\mathcal{N}=4$ supersymmetry.

The second example is of the t-flux tube closed to a torus with t-flux $e=1$ and vanishing $\beta$- and $\gamma$-flux
\be
\mathcal{I}_{g=1,m^-=0,\left(e=1,b=0,c=0\right)}^{N=2,k=2} & = & \kappa^{2}\oint\frac{dv_{1}}{4\pi iv_{1}}\oint\frac{dv_{2}}{4\pi iv_{2}}\frac{\Gamma_{e}\left(\frac{pq}{t}\left(\beta\gamma^{-1}\right)^{\pm1}v_{1}^{\pm1}v_{2}^{\pm1}\right)}{\Gamma_{e}\left(v_{1}^{\pm2}\right)\Gamma_{e}\left(v_{2}^{\pm2}\right)}\Gamma_{e}\left(pq\gamma^{-2}\beta^{-2}\right)\times\nonumber\\
 & & \Gamma_{e}\left(t\left(\gamma\beta^{-1}v_{2}\right)^{\pm1}v_{1}^{\pm1}\right)\Gamma_{e}\left(\left(\frac{pq}{t}\right)^{2}\right)\Gamma_{e}\left(\left(\frac{pq}{t}\right)^{2}\left(\gamma\beta^{-1}\right)^{\pm2}\right)\times\nonumber\\
 & & \Gamma_{e}\left(\left(\frac{pq}{t}\right)^{2}\left(\gamma\beta\right)^{2}\right)\kappa\oint\frac{dz}{4\pi iz}\frac{\Gamma_{e}\left(\frac{pq}{t\gamma\beta}\left(\beta\gamma^{-1}v_{2}^{-1}\right)^{\pm1}z^{\pm1}\right)}{\Gamma_{e}\left(z^{\pm2}\right)}\times\nonumber\\
 & & \Gamma_{e}\left(\gamma\beta z^{\pm1}v_{1}^{\pm1}\right)\mathcal{I}^{orbifold}_{\left\{ z_{1}=\beta\gamma^{-1},z_{2}=\frac{pq}{t}\right\} \left\{ c_{1}=v_{2},c_{2}=v_{1}\right\} ,\sqrt{zv_{2}},\sqrt{v_{2}/z}}\,.
\ee
In this case two $SU(2)$ gauge symmetries are Higgsed in the flow making several fields massive. Then, one $SU(2)$ gauge symmetry remains with only three flavors, and can be described in the IR by gauge invariants. The index of the IR theory is
\be
\mathcal{I}_{g=1,m^-=0,\left(e=1,b=0,c=0\right)}^{N=2,k=2,flow\ 1} & = & \Gamma_{e}\left(\left(\frac{pq}{t}\right)^{2}\epsilon^{\pm2}\right)^{2}\kappa\oint\frac{dv_{1}}{4\pi iv_{1}}\frac{\Gamma_{e}\left(\frac{pq}{t}\right)^{2}\Gamma_{e}\left(\frac{pq}{t}v_{1}^{\pm2}\right)}{\Gamma_{e}\left(v_{1}^{\pm2}\right)}\times\nonumber\\
 & & \kappa\oint\frac{dw_{1}}{4\pi iw_{1}}\frac{\Gamma_{e}\left(\frac{pq}{t}\right)^{2}\Gamma_{e}\left(\frac{pq}{t}w_{1}^{\pm2}\right)}{\Gamma_{e}\left(w_{1}^{\pm2}\right)}\Gamma_{e}\left(t^{1/2}\epsilon^{\pm1}w_{1}^{\pm1}v_{1}^{\pm1}\right)^{2}\nonumber\\
 & = & \mathcal{I}_{g=1,m^{-}=2,\left(e=1\right)}^{N=2,k=1}
\ee
with expected additional minimal punctures, flux and free chirals.

Initiating the second vev closes all minimal punctures and leaves us with the expected result
\be
\mathcal{I}_{g=1,m^-=0,\left(e=1,b=0,c=0\right)}^{N=2,k=2,flow\ 2} & = & \kappa\oint\frac{dv_{1}}{4\pi iv_{1}}\frac{\Gamma_{e}\left(\frac{pq}{t}\right)^{2}\Gamma_{e}\left(\frac{pq}{t}v_{1}^{\pm2}\right)}{\Gamma_{e}\left(v_{1}^{\pm2}\right)}\kappa\oint\frac{dw_{1}}{4\pi iw_{1}}\frac{\Gamma_{e}\left(\frac{pq}{t}\right)^{2}\Gamma_{e}\left(\frac{pq}{t}w_{1}^{\pm2}\right)}{\Gamma_{e}\left(w_{1}^{\pm2}\right)}\times\nonumber\\
 & & \Gamma_{e}\left(\left(\frac{pq}{t}\right)^{2}\right)^{2}\Gamma_{e}\left(t^{1/2}w_{1}^{\pm1}\right)^{3}\Gamma_{e}\left(\frac{t^{3/2}}{pq}w_{1}^{\pm1}\right)\Gamma_{e}\left(t^{1/2}w_{1}^{\pm1}v_{1}^{\pm2}\right)\nonumber\\
  & = & \mathcal{I}_{\left(e=2\right)}^{N=2,k=1}\,.
\ee

\subsubsection*{Genus $g=2$ Riemann surface}
The final examples we give are for genus 2 Riemann surfaces. These can be constructed by gluing the known indices of the interacting trinions of class $\mathcal{S}_2$ with $N=2$. The first example has no fluxes at all, given by
\be
\mathcal{I}_{g=2,m^{-}=0,\left(e=0,b=0,c=0\right)}^{N=2,k=2} & = & \Gamma_{e}\left(pq\left(\beta^{-1}\gamma\right)^{\pm2}\right)\kappa^{2}\oint\frac{du_{1}}{4\pi iu_{1}}\oint\frac{du_{2}}{4\pi iu_{2}}\frac{1}{\Gamma_{e}\left(u_{1}^{\pm2}\right)\Gamma_{e}\left(u_{2}^{\pm2}\right)}\times\nonumber\\
 & & \kappa^{2}\oint\frac{dc_{1}}{4\pi ic_{1}}\oint\frac{dc_{2}}{4\pi ic_{2}}\frac{1}{\Gamma_{e}\left(c_{1}^{\pm2}\right)\Gamma_{e}\left(c_{2}^{\pm2}\right)}\kappa\oint\frac{dy}{4\pi iy}\frac{1}{\Gamma_{e}\left(y^{\pm2}\right)}\times\nonumber\\
 & & \kappa^{2}\oint\frac{dv_{1}}{4\pi iv_{1}}\oint\frac{dv_{2}}{4\pi iv_{2}}\frac{1}{\Gamma_{e}\left(v_{1}^{\pm2}\right)\Gamma_{e}\left(v_{2}^{\pm2}\right)}\kappa\oint\frac{dz}{4\pi iz}\frac{1}{\Gamma_{e}\left(z^{\pm2}\right)}\times\nonumber\\
 & & \Gamma_{e}\left(\frac{pq\gamma}{t\beta}\left(\beta\gamma u_{2}^{-1}\right)^{\pm1}z^{\pm1}\right)\Gamma_{e}\left(\frac{t\beta}{\gamma}\left(\beta\gamma u_{2}^{-1}\right)^{\pm1}y^{\pm1}\right)\times\nonumber\\
 & & \Gamma_{e}\left(\frac{\beta}{\gamma}z^{\pm1}u_{1}^{\pm1}\right)\Gamma_{e}\left(\frac{\gamma}{\beta}y^{\pm1}u_{1}^{\pm1}\right)\mathcal{I}_{\mathbf{c},\mathbf{v},\sqrt{y/u_{2}},\left(u_{2}y\right)^{-1/2}}^{orbifold}\left(t\rightarrow\frac{pq}{t}\right)\times\nonumber\\
 & & \mathcal{I}_{\mathbf{c},\mathbf{v},\sqrt{zu_{2}},\sqrt{u_{2}/z}}^{orbifold}\left(\beta\rightarrow\beta^{-1},\gamma\rightarrow\gamma^{-1}\right)\,.
\ee
The vev that Initiates the flow causes one of the $SU(2)$ gauge symmetries (with fugacity $z$) to be Higgsed. Another $SU(2)$ gauge is left with only two flavors, which in the IR is described by the quantum deformed moduli space identifying two other $SU(2)$ gauge symmetries with one another. These processes give mass to several fields; thus, Higgsing two $SU(2)$ gauge symmetries of fugacities $c_1$ and $c_2$. In addition, two $SU(2)$ gauge symmetries of fugacities $v_1$ and $v_2$ are left with only three flavors; thus, described in the IR by quadratic gauge invariant composites. This complicated flow ends in the IR theory
\be
\mathcal{I}_{g=2,m^-=0,\left(e=0,b=0,c=0\right)}^{N=2,k=2,flow\ 1} & = & \kappa^{2}\oint\frac{dz_{1}}{4\pi iz_{1}}\oint\frac{dz_{2}}{4\pi iz_{2}}\frac{\Gamma_{e}\left(\sqrt{\frac{pq}{t}}\epsilon^{\pm1}z_{1}^{\pm1}z_{2}^{\pm1}\right)}{\Gamma_{e}\left(z_{1}^{\pm2}\right)\Gamma_{e}\left(z_{2}^{\pm2}\right)}\kappa\oint\frac{du_{2}}{4\pi iu_{2}}\times\nonumber\\
 & & \frac{\Gamma_{e}\left(\frac{pq}{t}\right)^{2}\Gamma_{e}\left(\frac{pq}{t}u_{2}^{\pm2}\right)}{\Gamma_{e}\left(u_{2}^{\pm2}\right)}\kappa^{2}\oint\frac{dy_{1}}{4\pi iy_{1}}\oint\frac{dy_{2}}{4\pi iy_{2}}\frac{\Gamma_{e}\left(\sqrt{\frac{pq}{t}}\epsilon^{\pm1}y_{1}^{\pm1}y_{2}^{\pm1}\right)}{\Gamma_{e}\left(y_{1}^{\pm2}\right)\Gamma_{e}\left(y_{2}^{\pm2}\right)}\times\nonumber\\
 & & \Gamma_{e}\left(\frac{pq}{t}\epsilon^{\pm2}\right)^{-1}\Gamma_{e}\left(t\right)^{2}\Gamma_{e}\left(t^{\frac{1}{2}}u_{2}^{\pm1}z_{1}^{\pm1}y_{2}^{\pm1}\right)\Gamma_{e}\left(t^{\frac{1}{2}}u_{2}^{\pm1}z_{2}^{\pm1}y_{1}^{\pm1}\right)\nonumber\\
 & = & \Gamma_{e}\left(\frac{pq}{t}\epsilon^{\pm2}\right)^{-1}\mathcal{I}_{g=2,m^{-}=2,\left(e=0\right)}^{N=2,k=1}\,,
\ee
with expected minimal punctures, flux and free chirals.

Initiating the second vev closes all minimal punctures, and on the way Higgses two $SU(2)$ gauge symmetries. The remaining IR theory is
\be
\mathcal{I}_{g=2,m^-=0,\left(e=0,b=0,c=0\right)}^{N=2,k=2,flow\ 2} & = & \kappa\oint\frac{du_{2}}{4\pi iu_{2}}\frac{\Gamma_{e}\left(\frac{pq}{t}u_{2}^{\pm2}\right)}{\Gamma_{e}\left(u_{2}^{\pm2}\right)}\kappa\oint\frac{dv_{2}}{4\pi iv_{2}}\frac{\Gamma_{e}\left(\frac{pq}{t}\right)^{2}\Gamma_{e}\left(\frac{pq}{t}v_{2}^{\pm2}\right)}{\Gamma_{e}\left(v_{2}^{\pm2}\right)}\times\nonumber\\
 & & \Gamma_{e}\left(\left(\frac{pq}{t}\right)^{2}\right)^{-1}\kappa\oint\frac{dz_{1}}{4\pi iz_{1}}\frac{\Gamma_{e}\left(\frac{pq}{t}\right)^{2}\Gamma_{e}\left(\frac{pq}{t}z_{1}^{\pm2}\right)}{\Gamma_{e}\left(z_{1}^{\pm2}\right)}\Gamma_{e}\left(t^{1/2}u_{2}^{\pm1}z_{1}^{\pm1}v_{2}^{\pm1}\right)^{2}\nonumber\\
 & = & \Gamma_{e}\left(\left(\frac{pq}{t}\right)^{2}\right)^{-1}\mathcal{I}_{g=2,m^-=0,\left(e=1\right)}^{N=2,k=1}\,,
\ee
again with expected free chirals and flux.

The second example is of t-flux $e=1$, $\beta$-flux $b=-1/2$ and $\gamma$-flux $c=1/2$
\be
\mathcal{I}_{g=2,m^{-}=0,\left(e=1,b=-1/2,c=1/2\right)}^{N=2,k=2} & = & \Gamma_{e}\left(pq\beta^{-2}\gamma^{2}\right)^{2}\kappa^{2}\oint\frac{du_{1}}{4\pi iu_{1}}\oint\frac{du_{2}}{4\pi iu_{2}}\frac{\Gamma_{e}\left(\frac{pq}{t}\left(\beta\gamma\right)^{\pm1}u_{1}^{\pm1}u_{2}^{\pm1}\right)}{\Gamma_{e}\left(u_{1}^{\pm2}\right)\Gamma_{e}\left(u_{2}^{\pm2}\right)}\times\nonumber\\
 & & \kappa^{2}\oint\frac{dc_{1}}{4\pi ic_{1}}\oint\frac{dc_{2}}{4\pi ic_{2}}\frac{\Gamma_{e}\left(\frac{pq}{t}\left(\beta\gamma^{-1}\right)^{\pm1}c_{1}^{\pm1}c_{2}^{\pm1}\right)}{\Gamma_{e}\left(c_{1}^{\pm2}\right)\Gamma_{e}\left(c_{2}^{\pm2}\right)}\times\nonumber\\
 & & \kappa^{2}\oint\frac{dv_{1}}{4\pi iv_{1}}\oint\frac{dv_{2}}{4\pi iv_{2}}\frac{\Gamma_{e}\left(\frac{pq}{t}\left(\beta\gamma^{-1}\right)^{\pm1}v_{1}^{\pm1}v_{2}^{\pm1}\right)}{\Gamma_{e}\left(v_{1}^{\pm2}\right)\Gamma_{e}\left(v_{2}^{\pm2}\right)}\times\nonumber\\
 & & \kappa\oint\frac{dz}{4\pi iz}\frac{\Gamma_{e}\left(\frac{pq\gamma}{t\beta}\left(\beta\gamma u_{2}^{-1}\right)^{\pm1}z^{\pm1}\right)}{\Gamma_{e}\left(z^{\pm2}\right)}\Gamma_{e}\left(t\left(\beta\gamma u_{2}^{-1}\right)^{\pm1}u_{1}^{\pm1}\right)\times\nonumber\\
 & & \kappa\oint\frac{dz}{4\pi iz}\frac{\Gamma_{e}\left(\frac{pq\gamma}{t\beta}\left(\beta\gamma u_{1}^{-1}\right)^{\pm1}y^{\pm1}\right)}{\Gamma_{e}\left(y^{\pm2}\right)}\Gamma_{e}\left(t\left(\beta\gamma u_{1}^{-1}\right)^{\pm1}u_{2}^{\pm1}\right)\times\nonumber\\
 & & \Gamma_{e}\left(\frac{\beta}{\gamma}z^{\pm1}u_{1}^{\pm1}\right)\mathcal{I}^{orbifold}_{\mathbf{c},\mathbf{v},\sqrt{zu_{2}},\sqrt{u_{2}/z}}(\beta\rightarrow\beta^{-1},\gamma\rightarrow\gamma^{-1})\times\nonumber\\
 & & \Gamma_{e}\left(\frac{\beta}{\gamma}y^{\pm1}u_{2}^{\pm1}\right)\mathcal{I}^{orbifold}_{\bar{\mathbf{c}},\bar{\mathbf{v}},\sqrt{yu_{1}},\sqrt{u_{1}/y}}(\beta\rightarrow\beta^{-1},\gamma\rightarrow\gamma^{-1})\,,
\ee
where the bars over the maximal puncture fugacities of the orbifold theory signals the switching of the two $SU(2)$ symmetries. This flow is simpler and after four $SU(2)$ gauge symmetries are Higgsed and several fields become massive we are left with
\be
\mathcal{I}_{g=2,m^{-}=0,\left(e=1,b=-1/2,c=1/2\right)}^{N=2,k=2,flow\ 1} & = & \kappa\oint\frac{dc_{1}}{4\pi ic_{1}}\frac{\Gamma_{e}\left(\frac{pq}{t}\right)^{2}\Gamma_{e}\left(\frac{pq}{t}c_{1}^{\pm2}\right)}{\Gamma_{e}\left(c_{1}^{\pm2}\right)}\kappa\oint\frac{dv_{1}}{4\pi iv_{1}}\frac{\Gamma_{e}\left(\frac{pq}{t}\right)^{2}\Gamma_{e}\left(\frac{pq}{t}v_{1}^{\pm2}\right)}{\Gamma_{e}\left(v_{1}^{\pm2}\right)}\times\nonumber\\
 & & \kappa^{2}\oint\frac{dz_{1}}{4\pi iz_{1}}\oint\frac{dz_{2}}{4\pi iz_{2}}\frac{\Gamma_{e}\left(\sqrt{\frac{pq}{t}}\epsilon^{\pm1}z_{1}^{\pm1}z_{2}^{\pm1}\right)}{\Gamma_{e}\left(z_{1}^{\pm2}\right)\Gamma_{e}\left(z_{2}^{\pm2}\right)}\Gamma_{e}\left(t^{\frac{1}{2}}\epsilon^{\pm1}z_{2}^{\pm1}v_{1}^{\pm1}\right)\times\nonumber\\
 & & \kappa^{2}\oint\frac{dy_{1}}{4\pi iy_{1}}\oint\frac{dy_{2}}{4\pi iy_{2}}\frac{\Gamma_{e}\left(\sqrt{\frac{pq}{t}}\epsilon^{\pm1}y_{1}^{\pm1}y_{2}^{\pm1}\right)}{\Gamma_{e}\left(y_{1}^{\pm2}\right)\Gamma_{e}\left(y_{2}^{\pm2}\right)}\Gamma_{e}\left(t^{\frac{1}{2}}\epsilon^{\pm1}y_{2}^{\pm1}v_{1}^{\pm1}\right)\times\nonumber\\
 & & \kappa^{2}\oint\frac{du_{1}}{4\pi iu_{1}}\oint\frac{du_{2}}{4\pi iu_{2}}\frac{\Gamma_{e}\left(\sqrt{\frac{pq}{t}}\epsilon^{\pm1}u_{1}^{\pm1}u_{2}^{\pm1}\right)}{\Gamma_{e}\left(u_{1}^{\pm2}\right)\Gamma_{e}\left(u_{2}^{\pm2}\right)}\times\nonumber\\
 & & \Gamma_{e}\left(t\right)\Gamma_{e}\left(t^{\frac{1}{2}}u_{2}^{\pm1}z_{1}^{\pm1}c_{1}^{\pm1}\right)\Gamma_{e}\left(t\right)\Gamma_{e}\left(t^{\frac{1}{2}}u_{1}^{\pm1}y_{1}^{\pm1}c_{1}^{\pm1}\right)\nonumber\\
 & = & \Gamma_{e}\left(\frac{pq}{t}\epsilon^{\pm2}\right)^{-2}\mathcal{I}_{g=2,m^{-}=5,\left(e=1/2\right)}^{N=2,k=1}\,,
\ee
with the expected minimal puncture, flux and free chirals.

Initiating the second vev closes all minimal punctures, and Higgses three additional $SU(2)$ gauge symmetries. The remaining IR theory is
\be
\mathcal{I}_{g=2,m^-=0,\left(e=1,b=-1/2,c=1/2\right)}^{N=2,k=2,flow\ 2} & = & \kappa\oint\frac{du_{1}}{4\pi iu_{1}}\frac{\Gamma_{e}\left(\frac{pq}{t}\right)^{2}\Gamma_{e}\left(\frac{pq}{t}u_{1}^{\pm2}\right)}{\Gamma_{e}\left(u_{1}^{\pm2}\right)}\kappa\oint\frac{dc_{1}}{4\pi ic_{1}}\frac{\Gamma_{e}\left(\frac{pq}{t}\right)^{2}\Gamma_{e}\left(\frac{pq}{t}c_{1}^{\pm2}\right)}{\Gamma_{e}\left(c_{1}^{\pm2}\right)}\times\nonumber\\
 & & \kappa\oint\frac{dv_{1}}{4\pi iv_{1}}\frac{\Gamma_{e}\left(\frac{pq}{t}\right)^{2}\Gamma_{e}\left(\frac{pq}{t}v_{1}^{\pm2}\right)}{\Gamma_{e}\left(v_{1}^{\pm2}\right)}\kappa\oint\frac{dz_{1}}{4\pi iz_{1}}\frac{\Gamma_{e}\left(\frac{pq}{t}\right)^{2}\Gamma_{e}\left(\frac{pq}{t}z_{1}^{\pm2}\right)}{\Gamma_{e}\left(z_{1}^{\pm2}\right)}\times\nonumber\\
 & & \kappa\oint\frac{dy_{1}}{4\pi iy_{1}}\frac{\Gamma_{e}\left(\frac{pq}{t}y_{1}^{\pm2}\right)}{\Gamma_{e}\left(y_{1}^{\pm2}\right)}\Gamma_{e}\left(\frac{t}{\sqrt{pq}}z_{1}^{\pm1}c_{1}^{\pm1}\right)\Gamma_{e}\left(t^{1/2}u_{1}^{\pm1}z_{1}^{\pm1}v_{1}^{\pm1}\right)\times\nonumber\\
 & & \Gamma_{e}\left(\frac{t}{\sqrt{pq}}y_{1}^{\pm1}c_{1}^{\pm1}\right)\Gamma_{e}\left(t^{1/2}u_{1}^{\pm1}y_{1}^{\pm1}v_{1}^{\pm1}\right)\nonumber\\
 & = & \Gamma_{e}\left(\left(\frac{pq}{t}\right)^{2}\right)^{-2}\mathcal{I}_{g=2\left(e=3\right)}^{N=2,k=1}\,,
\ee
with the expected free chirals and flux.

In Appendix \ref{A:Flows punc}, we show additional explicit flow examples of Lagrangian theories with punctures in class $\mathcal{S}_2$ with $N=2$.

\subsection{Flows from $\mathcal{S}_k$ to $\mathcal{S}_1$ with general $N$}
In the general case of class $\mathcal{S}_k$ and general $N$ the only known punctureless Lagrangians are of tori theories with zero t-flux (see \cite{Bah:2017gph}). These theories can be built by starting with a chain of free trinions glued to a torus and closing all remaining minimal punctures. For this model we will only focus on the case where we give vevs that break all the $\beta_i$ and $\gamma_j$ internal symmetries. In the other cases the results are as expected but too cluttered and complicated to be presented. Therefore, as stated above we will only pursue the case of maximal internal symmetries breaking, with baryon vevs taking the form
\be
\left(\beta_{i}^{-1}\gamma_{j}\left(\frac{pq}{t}\right)\right)^{N}=1\,,
\ee
with $i=1,...,k-1$ and $j=1,...,k$ translating to the assignments $\beta_{i}=\frac{pq}{t},\,\gamma_{j}=1$.

We choose an example of a torus that will contain all the required poles with fluxes $\mathcal{F}=\left(e=0,b=\left(b_{i\ne k}=-1,b_{k}=k-1\right),c=\vec{0}\right)$, and index
\be
\mathcal{I}_{g=1,\left(e=0,b=\left(b_{i\ne k}=-1,b_{k}=k-1\right),c=\vec{0}\right)}^{N,k} & = & \prod_{a=1}^{k-1}\Gamma_{e}\left(pq\beta_{k}^{N}\beta_{a}^{-N}\right)^{k}\left(\frac{\kappa^{N-1}}{N!}\right)^{k(k-1)}\prod_{a+b\ne k}^{k}\prod_{i=1}^{N-1}\oint\frac{dv_{a,b}^{(i)}}{2\pi iv_{a,b}^{(i)}}\times\nonumber\\
 & & \frac{\prod_{a+b\ne k,k-1}^{k}\prod_{i,j=1}^{N}\Gamma_{e}\left(\frac{pq}{t}\beta_{a+b}^{-1}\gamma_{a+1}v_{a,b}^{(i)}\left(v_{a+1,b}^{(j)}\right)^{-1}\right)}{\prod_{a+b\ne k}^{k}\prod_{i\ne j}^{N}\Gamma_{e}\left(v_{a,b}^{(i)}\left(v_{a,b}^{(j)}\right)^{-1}\right)}\times\nonumber\\
 & & \prod_{a=1}^{k}\prod_{i,j=1}^{N}\Gamma_{e}\left(\frac{pq}{t}\beta_{k-1}^{-1}\gamma_{a+1}v_{a,k-a-1}^{(i)}\left(v_{a+1,k-a}^{(j)}\right)^{-1}\right)\times\nonumber\\
 & & \prod_{a=1}^{k}\prod_{i,j=1}^{N}\Gamma_{e}\left(\beta_{k-1}\beta_{k}^{-1}\left(v_{a,k-1-a}^{(i)}\right)^{-1}v_{a,k-a+1}^{(j)}\right)\times\nonumber\\
 & & \prod_{a+b\ne k,k-1}^{k}\prod_{i,j=1}^{N}\Gamma_{e}\left(\beta_{a+b}\beta_{k}^{-1}\left(v_{a,b}^{(i)}\right)^{-1}v_{a,b+1}^{(j)}\right)\times\nonumber\\
 & & \prod_{a=1}^{k}\prod_{i,j=1}^{N}\Gamma_{e}\left(t\gamma_{a+1}^{-1}\beta_{k}v_{a+1,k-a}^{(i)}\left(v_{a,k-a+1}^{(j)}\right)^{-1}\right)\times\nonumber\\
 & & \prod_{a+b\ne k,k-1}^{k}\prod_{i,j=1}^{N}\Gamma_{e}\left(t\gamma_{a+1}^{-1}\beta_{k}v_{a+1,b}^{(i)}\left(v_{a,b+1}^{(j)}\right)^{-1}\right)\,,
\ee
where $v_{a,b}^{(i)}$ are the fugacities of the $k(k-1)$ $SU(N)$ gauge symmetries, with $a,b=1,...,k$ going over the different gauge factors, and $i=1,...,N$ running over the $N$ fugacities of each gauge group (with product of all equal $1$). In addition, $a$ and $b$ are defined cyclically, meaning we always consider $a\rightarrow (a-1 \mod k)+1$, such that they are always given by an integer number between $1$ and $k$. The flow is pretty straight forward, with $k^2-k-1$ of the gauge $SU(N)$ symmetries Higgsed and subsequently many fields becoming massive. The last $SU(N)$ gauge symmetry is removed since all remaining multiplets are massive and decouple in the IR. Finally we use the charge redefinition $\left(\frac{pq}{t}\right)\rightarrow \left(\frac{pq}{t}\right)^{1/k}$ and find
\be
\mathcal{I}_{g=1,\left(e=0,b=\left(b_{i\ne k}=-1,b_{k}=k-1\right),c=\vec{0}\right)}^{N,k,flow} & = & \Gamma_{e}\left(\left(\frac{pq}{t}\right)^{N}\right)^{-k(k-1)}\left(\frac{\kappa^{N-1}}{N!}\right)\nonumber\\
 & = & \Gamma_{e}\left(\left(\frac{pq}{t}\right)^{N}\right)^{-k(k-1)}\mathcal{I}_{g=1,\left(e=0\right)}^{N,k=1}
\ee
We find the expected flux from equation \eqref{E:tFluxFlow} is $e_1=k\cdot 0+(k-1)(1-1)=0$ and the expected free chirals from equation \eqref{E:freeChirals} $n_{FC}=-k(k-1)-(k-1)(0-0)=-k(k-1)$. In addition as was found before in the case of zero t-flux torus; the IR theory has enhanced supersymmetry of $\mathcal{N}=4$.  

\section{$4d$ Anomaly polynomial flow}
\label{S:APFlow}
In this section we will analyze the flow on the four dimensional anomaly polynomial. Since this anomaly polynomial can be inferred directly from the six dimensional anomaly polynomial only on a closed Riemann surface with no punctures, we will consider the flow generated by baryon vevs $\left(\beta_{n}^{-1}\gamma_{m}\left(\frac{pq}{t}\right)\right)^{N}=1$ and $\left(\beta_{n}^{-1}\gamma_{k}\left(\frac{pq}{t}\right)\right)^{N}=1$, with $n,m=1,..,\ell$. This will allow for easy comparison as we do not generate any new minimal punctures in this flow.

Starting from the anomaly polynomial 8-form of the $6d$ theory described by $N$ M5-branes probing a $\mathbb{C}^2/\mathbb{Z}_k$ singularity, one can compactify it on a closed Riemann surface and find the $4d$ anomaly polynomial 6-form. This was done in \cite{Bah:2017gph}, and we reproduce the result here in our conventions
\be
I_{6} & = & -\frac{(k^{2}-2)(N-1)}{2}(2g-2)c_{1}(R')-k^{2}NN_{e}^{(k)}c_{1}(t)\nonumber\\
 & & +\frac{(N-1)(k^{2}(N^{2}+N-1)+2)}{12}(2g-2)c_{1}(R')^{3}+\frac{k^{2}N(N^{2}-1)}{6}N_{e}^{(k)}c_{1}(R')^{2}c_{1}(t)\nonumber\\
 & & -\frac{k^{2}N(N^{2}-1)}{12}(2g-2)c_{1}(R')c_{1}(t)^{2}-\frac{k^{2}N^{3}}{6}N_{e}^{(k)}c_{1}(t)^{3}\nonumber\\
 & & +\frac{1}{2}\left(-kN(N-1)c_{1}(R')^{2}+kN^{2}c_{1}(t)^{2}\right)\sum_{i=1}^{k}\left(N_{b_{i}}^{(k)}c_{1}(\beta_{i})+N_{c_{i}}^{(k)}c_{1}(\gamma_{i})\right)\nonumber\\
 & & -\frac{kN^{2}(N-1)}{4}(2g-2)c_{1}(R')\sum_{i=1}^{k}\left(c_{1}(\beta_{i})^{2}+c_{1}(\gamma_{i})^{2}\right)+kN\sum_{i=1}^{k}\left(N_{b_{i}}^{(k)}c_{1}(\beta_{i})+N_{c_{i}}^{(k)}c_{1}(\gamma_{i})\right)\nonumber\\
 & & -\frac{kN^{2}}{2}\sum_{i=1}^{k}\left(\left(NN_{e}^{(k)}-N_{b_{i}}^{(k)}\right)c_{1}(t)c_{1}(\beta_{i})^{2}+\left(NN_{e}^{(k)}+N_{c_{i}}^{(k)}\right)c_{1}(t)c_{1}(\gamma_{i})^{2}\right)\nonumber\\
 & & +\frac{N^{2}(N-1)}{2}\left(\left(\sum_{i=1}^{k}c_{1}(\beta_{i})^{2}\right)\left(\sum_{j=1}^{k}N_{b_{j}}^{(k)}c_{1}(\beta_{j})\right)+\left(\sum_{i=1}^{k}c_{1}(\gamma_{i})^{2}\right)\left(\sum_{j=1}^{k}N_{c_{j}}^{(k)}c_{1}(\gamma_{j})\right)\right)\nonumber\\
 & & +\frac{kN^{3}}{6}\sum_{i=1}^{k}\left(\left(N_{b_{i}}^{(k)}-N_{e}^{(k)}\right)c_{1}(\beta_{i})^{3}+\left(N_{c_{i}}^{(k)}+N_{e}^{(k)}\right)c_{1}(\gamma_{i})^{3}\right)\nonumber\\
 & & +\frac{N^{2}}{2}\left(\left(\sum_{i=1}^{k}c_{1}(\beta_{i})^{2}\right)\left(\sum_{j=1}^{k}N_{c_{j}}^{(k)}c_{1}(\gamma_{j})\right)+\left(\sum_{i=1}^{k}c_{1}(\gamma_{i})^{2}\right)\left(\sum_{j=1}^{k}N_{b_{j}}^{(k)}c_{1}(\beta_{j})\right)\right)\,,
\ee
where $c_1(X)$ is the first Chern class of $U(1)_X$, and $N_e^{(k)}, N_{b_i}^{(k)}, N_{c_j}^{(k)}$ are the same as $e^{(k)}, b_i^{(k)},c_j^{(k)}$ from before, meaning the t- $\beta_i$- and $\gamma_j$-flux (the change in notation was done to avoid confusion). $R'$ is the R-symmetry arising naturally from $6d$, which is different from the $4d$ R-symmetry we used before, and both are not necessarily the conformal R-symmetry. The relation between the two is via $Q_{R'}=Q_R+Q_t$, where $Q_X$ is the charge of a field under $U(1)_X$.

We can use the vevs we discussed before \eqref{E:General vevs} to initiate the flow from class $\mathcal{S}_k$ to $\mathcal{S}_{k-\ell}$ with no additional punctures on the level of the anomaly polynomial as another independent check of our results. This is done through several stages, first we need to change to the R-symmetry conventions we use in $4d$. This is done by the assignments
\be
c_{1}\left(t\right)=c_{1}\left(t\right)-c_{1}\left(R\right)\,,\quad c_{1}\left(R'\right)=c_{1}\left(R\right)\,.
\ee
The next stage is to give vevs using the following assignments
\be
c_{1}\left(\beta_{n}\right) & = & 2c_{1}\left(R\right)-c_{1}\left(t\right)+\frac{k-\ell}{k}c_{1}\left(\widetilde{\gamma}_{k-\ell}\right)\,,\qquad n=1,...,\ell\nonumber\\
c_{1}\left(\beta_{i}\right) & = & c_{1}\left(\widetilde{\beta}_{i-\ell}\right)-\frac{\ell}{k-\ell}\left(2c_{1}\left(R\right)-c_{1}\left(t\right)\right)-\frac{\ell}{k}c_{1}\left(\widetilde{\gamma}_{k-\ell}\right)\,,\qquad i=\ell+1,...,k\nonumber\\
c_{1}\left(\gamma_{m}\right) & = & \frac{k-\ell}{k}c_{1}\left(\widetilde{\gamma}_{k-\ell}\right)\,,\qquad m=1,...,\ell\nonumber\\
c_{1}\left(\gamma_{j}\right) & = & c_{1}\left(\widetilde{\gamma}_{j-\ell}\right)-\frac{\ell}{k}c_{1}\left(\widetilde{\gamma}_{k-\ell}\right)\,,\qquad j=\ell+1,...,k\,.
\ee
Followed by the conventional t-charge and R-charge redefinition, produced by
\be
c_{1}\left(t\right)=\frac{k-\ell}{k}c_{1}\left(t\right)+\frac{2\ell}{k}c_{1}\left(R\right)\,.
\ee
Before moving back to the $6d$ R-symmetry conventions, we need to remove the free chiral fields that decouple in the flow. Therefore, the free chiral fields contribution to the anomaly polynomial 6-form need to be deducted. The contribution is
\be
I_{6}^{FC} & = & \sum_{i=1}^{k-\ell}\left(\sum_{n=1}^{\ell}N_{b_{n}}^{(k)}-\ell N_{b_{i+\ell}}^{(k)}+\ell\left(g-1\right)\right)\times\nonumber\\
 & & \qquad\left(\frac{1}{6}\left(-Nc_{1}\left(\widetilde{\beta}_{i}\right)+Nc_{1}\left(\widetilde{\gamma}_{k-\ell}\right)+(2N-1)c_{1}\left(R\right)-Nc_{1}\left(t\right)\right){}^{3}\right.\nonumber\\
 & & \qquad\quad\left.+\left(-Nc_{1}\left(\widetilde{\beta}_{i}\right)+Nc_{1}\left(\widetilde{\gamma}_{k-\ell}\right)+(2N-1)c_{1}\left(R\right)-Nc_{1}\left(t\right)\right)\right)\nonumber\\
 & & +\sum_{i=1}^{k-\ell-1}\left(-\sum_{n=1}^{\ell}N_{b_{n}}^{(k)}+\ell\left(N_{c_{j+\ell}}^{(k)}+NN_{e}^{(k)}+\left(N-1\right)\left(g-1\right)\right)\right)\times\nonumber\\
 & & \qquad\left(\frac{1}{6}\left(Nc_{1}\left(\widetilde{\gamma}_{i}\right)-Nc_{1}\left(\widetilde{\gamma}_{k-\ell}\right)-c_{1}\left(R\right)\right){}^{3}+\left(Nc_{1}\left(\widetilde{\gamma}_{i}\right)-Nc_{1}\left(\widetilde{\gamma}_{k-\ell}\right)-c_{1}\left(R\right)\right)\right)\nonumber\\
 & & +\left(\sum_{i=1}^{\ell}\left(-\left(\ell+1\right)N_{b_{i}}^{(k)}+\ell N_{c_{i}}^{(k)}\right)+\ell N_{c_{k}}^{(k)}+\ell\left(\ell+1\right)\left(NN_{e}^{(k)}+\left(N-1\right)\left(g-1\right)\right)\right)\times\nonumber\\
 & & \qquad\left(\frac{1}{6}(-c_{1}\left(R\right))^{3}-c_{1}\left(R\right)\right)\,,
\ee
where the last two lines are the contribution from the divergence related to Higgs bosons that we need to remove as well.

Finally we return to the $6d$ R-symmetry conventions by the assignments
\be
c_{1}\left(t\right)=c_{1}\left(t\right)+c_{1}\left(R'\right)\,,\quad c_{1}\left(R\right)=c_{1}\left(R'\right)\,,
\ee
and the resulting anomaly polynomial 6-form is
\be
I_{6}^{flow} & = & -\frac{(\left(k-\ell\right)^{2}-2)(N-1)}{2}(2g-2)c_{1}(R')-\left(k-\ell\right)^{2}NN_{e}^{(k-\ell)}c_{1}(t)\nonumber\\
 & & +\frac{(N-1)(\left(k-\ell\right)^{2}(N^{2}+N-1)+2)}{12}(2g-2)c_{1}(R')^{3}\nonumber\\
 & & +\frac{\left(k-\ell\right)^{2}N(N^{2}-1)}{6}N_{e}^{(k-\ell)}c_{1}(R')^{2}c_{1}(t)\nonumber\\
 & & -\frac{\left(k-\ell\right)^{2}N(N^{2}-1)}{12}(2g-2)c_{1}(R')c_{1}(t)^{2}-\frac{\left(k-\ell\right)^{2}N^{3}}{6}N_{e}^{(k-\ell)}c_{1}(t)^{3}\nonumber\\
 & & +\frac{k-\ell}{2}\left(N(N-1)c_{1}(R')^{2}+N^{2}c_{1}(t)^{2}\right)\sum_{i=1}^{k-\ell}\left(N_{b_{i}}^{(k-\ell)}c_{1}\left(\widetilde{\beta}_{i}\right)+N_{c_{i}}^{(k-\ell)}c_{1}\left(\widetilde{\gamma}_{i}\right)\right)\nonumber\\
 & & -\frac{\left(k-\ell\right)N^{2}(N-1)}{4}(2g-2)c_{1}(R')\sum_{i=1}^{k-\ell}\left(c_{1}\left(\widetilde{\beta}_{i}\right)^{2}+c_{1}\left(\widetilde{\gamma}_{i}\right)^{2}\right)\nonumber\\
 & & +\left(k-\ell\right)N\sum_{i=1}^{k-\ell}\left(N_{b_{i}}^{(k)}c_{1}\left(\widetilde{\beta}_{i}\right)+N_{c_{i}}^{(k)}c_{1}\left(\widetilde{\gamma}_{i}\right)\right)\nonumber\\
 & & -\frac{\left(k-\ell\right)N^{2}}{2}\sum_{i=1}^{k-\ell}\left(NN_{e}^{(k-\ell)}-N_{b_{i}}^{(k-\ell)}\right)c_{1}(t)c_{1}\left(\widetilde{\beta}_{i}\right)^{2}\nonumber\\
 & & -\frac{\left(k-\ell\right)N^{2}}{2}\sum_{i=1}^{k-\ell}\left(NN_{e}^{(k-\ell)}+N_{c_{i}}^{(k-\ell)}\right)c_{1}(t)c_{1}\left(\widetilde{\gamma}_{i}\right)^{2}\nonumber\\
 & & +\frac{N^{2}(N-1)}{2}\left(\sum_{i=1}^{k-\ell}c_{1}\left(\widetilde{\beta}_{i}\right)^{2}\right)\left(\sum_{j=1}^{k-\ell}N_{b_{j}}^{(k-\ell)}c_{1}\left(\widetilde{\beta}_{j}\right)\right)\nonumber\\
 & & +\frac{N^{2}(N-1)}{2}\left(\sum_{i=1}^{k-\ell}c_{1}\left(\widetilde{\gamma}_{i}\right)^{2}\right)\left(\sum_{j=1}^{k-\ell}N_{c_{j}}^{(k-\ell)}c_{1}\left(\widetilde{\gamma}_{j}\right)\right)\nonumber\\
 & & +\frac{\left(k-\ell\right)N^{3}}{6}\sum_{i=1}^{k-\ell}\left(\left(N_{b_{i}}^{(k-\ell)}-N_{e}^{(k-\ell)}\right)c_{1}\left(\widetilde{\beta}_{i}\right)^{3}+\left(N_{c_{i}}^{(k-\ell)}+N_{e}^{(k-\ell)}\right)c_{1}\left(\widetilde{\gamma}_{i}\right)^{3}\right)\nonumber\\
 & & +\frac{N^{2}}{2}\left(\sum_{i=1}^{k-\ell}c_{1}\left(\widetilde{\beta}_{i}\right)^{2}\right)\left(\sum_{j=1}^{k-\ell}N_{c_{j}}^{(k-\ell)}c_{1}\left(\widetilde{\gamma}_{j}\right)\right)\nonumber\\
 & & +\frac{N^{2}}{2}\left(\sum_{i=1}^{k-\ell}c_{1}\left(\widetilde{\gamma}_{i}\right)^{2}\right)\left(\sum_{j=1}^{k-\ell}N_{b_{j}}^{(k-\ell)}c_{1}\left(\widetilde{\beta}_{j}\right)\right)\,,
\ee
where the fluxes are as we expected
\be
N_{e}^{(k-\ell)} & = & \frac{k}{k-\ell}N_{e}^{(k)}+\frac{\ell}{k-\ell}\left(g-1\right)\nonumber\\
N_{b_{i}}^{(k-\ell)} & = & N_{b_{i+\ell}}^{(k)}+\frac{1}{k-\ell}\sum_{n=1}^{\ell}N_{b_{n}}^{(k)}\nonumber\\
N_{c_{j}}^{(k-\ell)} & = & N_{c_{j+\ell}}^{(k)}+\frac{1}{k-\ell}\left(\sum_{i=1}^{\ell}N_{b_{i}}^{(k)}-N\ell N_{e}^{(k)}-N\ell\left(g-1\right)\right)\,.
\ee
This is a highly non trivial test for our predictions from the Coulomb limit formula given in \eqref{E:tFluxGen} and \eqref{E:bcFluxGen}, especially due to the nonlinear anomalies matching.

\section{Discussion}
In this paper we have studied RG flows starting with the $6d$ $\mathcal{T}(SU(k),N)$ $(1,0)$ SCFT and ending in a  $A_{N-1}$ type class $\mathcal{S}_{k'}$ theory. We compared two ways to generate such a flow, the $6d\rightarrow 6d \rightarrow 4d$ and the $6d\rightarrow 4d \rightarrow 4d$. The former defined by first giving a $6d$ vev to the so called end to end operator generating an RG flow to $\mathcal{T}(SU(k'),N)$ with $k'<k$, and then compactifying to the class $\mathcal{S}_{k'}$ theory. The latter defined by first compactifying to class $\mathcal{S}_k$ theory and then generating a flow by a $4d$ vev to the same class $\mathcal{S}_{k'}$ theory.

We found that the two flow sequences can match when we choose both the $6d$ and $4d$ vevs to be constant, with the $4d$ vev given to an operator which is a natural reduction of the $6d$ operator that the $6d$ vev is given to. We found that in the general case the matching will be done with compactifications on Riemann surfaces differing by the number of punctures. 
We have argued that the number of punctures in the two sequences is different, since a proper way to define the flow is by turning on both deformations, vev and compact geometry, simultaneously. Due to the flux, this leads to a non constant value for the vev.
We have also derived how the fluxes on the Riemann surface match between the two flows. The matching of the two types of  flows was tested with a  variety of tools. These include the $6d$ and $4d$ anomaly polynomials, class $\mathcal{S}_k$ Coulomb index, and the full superconformal index. 

On one hand our results can be viewed as a farther test of the map between geometric engineering of $4d$ theories starting with a compactification of $6d$ SCFT and concrete four dimensional constructions. 
On the other hand, the fact that $6d$ flows involving vevs and compact geometry when flux is turned on can be connected to compactifications with punctures might have more general applications.  For example, although some compactifications with flux and number of punctures less or equal to two are understood starting with a wider class of $6d$ theories (see {\it e.g.} \cite{Kim:2018lfo}), understanding of surfaces with higher number of punctures is lacking. Thus flows generating punctures might turn out to be of use for these classes of theories, expanding our understanding of the geometric constructions of four dimensional SCFTs.

\section*{Acknowledgments}

The research of SSR and ES is supported by Israel Science Foundation under grant no. 2289/18 and by I-CORE  Program of the Planning and Budgeting Committee. GZ is supported in part by  World Premier International Research Center Initiative (WPI), MEXT, Japan.

\vspace{10pt}
\begin{appendix}
\vspace{10pt}

\section{Class $\mathcal{S}_k$ puncture conventions}
\label{A:PuncConventions}
\begin{figure}[t]
	\centering
  	\includegraphics[scale=0.35]{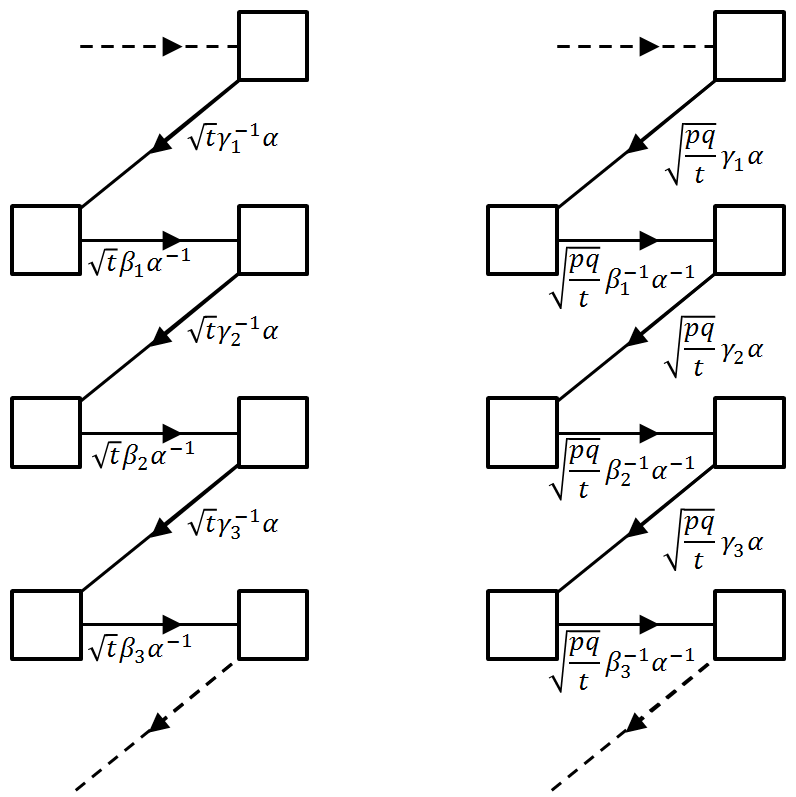}
    \caption{A generalized quiver description of class $\mathcal{S}_k$ free trinion. The squares represent flavor $SU(N)$ symmetries and there are $k$ of them on each side of each diagram winding in a manner that connects the upper and lower parts of each diagram. The arrows represent chiral multiplets in the bifundamental representation of the two corresponding $SU(N)$ flavor symmetries. \textbf{On the left:} The free trinion theory with $4d$ R-charge conventions giving all shown chiral fields R-charge 0. The internal symmetries $U(1)^{k}_{\beta} \times U(1)^{k}_{\gamma} \times U(1)_t$ coming from the $6d$ global symmetries $SU(N)_{\beta} \times SU(N)_{\gamma} \times U(1)_t$ are denoted accordingly by their fugacities. In addition the fugacity $\alpha$ is related to the $U(1)$ flavor symmetry associated to the free trinion minimal puncture. \textbf{On the right:} The negative free trinion with $\beta,\,\gamma$ and $t$ charges flipped and R-charge shifted by $Q_R\rightarrow 2Q_t-Q_R$. This procedure flips the sign of all puncture. The color of maximal punctures is related to the coupling of $\beta$ and $\gamma$ charges on the chiral fields. In the cases above we define the left punctures as having color $1$ and the right as having color $2$.}
    \label{F:FreeTrinions}
\end{figure}

In this appendix we will shortly review the puncture conventions of class $\mathcal{S}_k$. For a more extensive description one may look at \cite{Gaiotto:2015usa}. To that end we must first discuss the basic building block of class $\mathcal{S}_k$, the free trinion. This trinion is described by a sphere with two maximal punctures each associated with an $SU(N)^k$ flavor symmetry, and one minimal puncture associated to a $U(1)$ flavor symmetry. The free trinion model can be described in a generalized quiver description as shown on the left side of Fig. \ref{F:FreeTrinions}.

Maximal punctures have a few properties separating them from one another. One such property is the color depending on the pairing of $\beta_i$-s and $\gamma_j$-s of the chiral fields charged under couples of $SU(N)$-s of the puncture. In addition, one can consider the sign property of both the minimal and maximal punctures. In the case of the free trinion as shown on the left of Fig. \ref{F:FreeTrinions} all punctures are of positive sign, but by flipping the charges as shown on the right of Fig. \ref{F:FreeTrinions} one can get the negative free trinion with all punctures being of negative sign. One can also flip the sign of a maximal puncture by adding bifundamental chiral multiplets transforming under pairs of $SU(N)$ symmetries of the same puncture. Such chirals need to be coupled through the superpotential to the mesonic operators associated to the maximal puncture. These chirals are the same ones added in the process of $\Phi$-gluing.

In general one can glue two maximal punctures. We consider two such gluings:
\begin{itemize}
\item $\Phi$-gluing - Gluing two maximal punctures with the same sign and color. The gluing procedure requires adding an ${\cal N}=1$ $SU(N)^k$ vector multiplet, as well as bifundamental chiral multiplets between each two $SU(N)$ gauge groups along the gluing. These bifundamentals are coupled through the superpotential to the mesonic operators associated to the maximal punctures glued.
\item $S$-gluing - Gluing two maximal punctures with an opposite sign and same color. The gluing requires only adding an ${\cal N}=1$ $SU(N)^k$ vector multiplet, and coupling mesonic operators associated to both maximal punctures through the superpotential.
\end{itemize}

There are various other punctures besides minimal and maximal that can be found by giving vev to operators charged under maximal punctures and partially closing them, but these will not be needed for the scope of this manuscript. One deformation that will proof useful is the closure of minimal punctures. This is achieved by giving a vev to a baryon charged under the minimal puncture $U(1)$ symmetry.

\section{$\mathcal{N}=1$ superconformal index}\label{A:indexdefinitions}
In this appendix we give a short introduction of the $\mathcal{N}=1$ superconformal index \cite{Kinney:2005ej,Romelsberger:2005eg}, some related notations, and usful results. For more comprehensive explanations and definitions see \cite{Rastelli:2016tbz}.
The index of an SCFT is defined as the Witten index of the theory in radial quantization. The index in four dimensions is defined as a trace over the Hilbert space of the theory quantized on $\mathbb{S}^3$
\be
\mathcal{I}\left(\mu_i\right)=Tr(-1)^F e^{-\beta \delta} e^{-\mu_i \mathcal{M}_i},
\ee
where $\delta\triangleq \half \left\{\mathcal{Q},\mathcal{Q}^{\dagger}\right\}$, with $\mathcal{Q}$ one of the Poincar\'e supercharges, and $\mathcal{Q}^{\dagger}=\mathcal{S}$ it's conjugate conformal supercharge, $\mathcal{M}_i$ are $\mathcal{Q}$-closed conserved charges and $\mu_i$ their associated chemical potentials. All the contributing states are with $\delta=0$ making the index independent on $\beta$, since states with $\delta>0$ come in boson/fermion pairs with opposite contributions.

For $\mathcal{N}=1$, the supercharges are $\left\{\mathcal{Q}_{\alpha},\,\mathcal{S}^{\alpha} \triangleq \mathcal{Q}^{\dagger\alpha},\,\widetilde{\mathcal{Q}}_{\dot{\alpha}},\,\widetilde{\mathcal{S}}^{\dot{\alpha}} \triangleq \widetilde{\mathcal{Q}}^{\dagger\dot{\alpha}}\right\}$, with $\alpha=\pm$ and $\dot{\alpha}=\dot{\pm}$ the respective $SU(2)_1$ and $SU(2)_2$ indices of the isometry group of $\mathbb{S}^3$ ($Spin(4)=SU(2)_1 \times SU(2)_2$).
Since different choices of $\mathcal{Q}$ in the definition of the index lead to physically equivalent indices, we simply choose $\mathcal{Q}=\widetilde{\mathcal{Q}}_{\dot{-}}$. Under this choice the index trace formula takes the form
\be
\mathcal{I}\left(p,q\right)=Tr(-1)^F p^{j_1 + j_2 +\half r} q^{j_2 - j_1 +\half r}.
\ee
where $p$ and $q$ are fugacities associated with the supersymmetry preserving squashing of the $\mathbb{S}^3$ \cite{Dolan:2008qi}. $j_1$ and $j_2$ are the Cartan generators of $SU(2)_1$ and $SU(2)_2$, and $r$ is the generator of the $U(1)_r$ R-symmetry.

The index is computed by listing all gauge invariant operators one can construct from modes of the fields. The modes and operators are conventionally called "letters" and "words", respectively. The single-letter index for a vector multiplet and a chiral multiplet transforming in the $\mathcal{R}$ representation of the gauge$\times$flavor group is
\be
i_V \left(p,q,U\right) & = & \frac{2pq-p-q}{(1-p)(1-q)} \chi_{adj}\left(U\right), \nonumber\\
i_{\chi(r)}\left(p,q,U,V\right) & = & 
\frac{(pq)^{\half r} \chi_{\mathcal{R}} \left(U,V\right) - (pq)^{\frac{2-r}{2}} \chi_{\bar{\mathcal{R}}} \left(U,V\right)}{(1-p)(1-q)},
\ee
where $\chi_{\mathcal{R}} \left(U,V\right)$ and $\chi_{\bar{\mathcal{R}}} \left(U,V\right)$ denote the characters of $\mathcal{R}$ and the conjugate representation $\bar{\mathcal{R}}$, with $U$ and $V$ gauge and flavor group matrices, respectively.

With the single letter indices at hand, we can write the full index by listing all the words and projecting them to gauge singlets by integrating over the Haar measure of the gauge group. This takes the general form
\be
\mathcal{I} \left(p,q,V\right)=\int \left[dU\right] \prod_{k} PE\left[i_k\left(p,q,U,V\right)\right],
\ee
where $k$ labels the different multiplets in the theory, and $PE[i_k]$ is the plethystic exponent of the single-letter index of the $k$-th multiplet, responsible for listing all the words. The plethystic exponent is defined by
\be
PE\left[i_k\left(p,q,U,V\right)\right] \triangleq \exp \left\{ \sum_{n=1}^{\infty} \frac{1}{n} i_k\left(p^n,q^n,U^n,V^n\right) \right\}.
\ee

Specializing to the case of $SU(N_c)$ gauge group. The full contribution for a chiral superfield in the fundamental representation of $SU(N_c)$ with R-charge $r$ can be written in terms of elliptic gamma functions, as follows
\be
PE\left[i_k\left(p,q,U\right)\right] & \equiv & \prod_{i=1}^{N_c} \Gamma_e \left((pq)^{\half r} z_i \right), \nonumber \\
\Gamma_e(z)\triangleq\Gamma\left(z;p,q\right) & \equiv & \prod_{n,m=0}^{\infty} \frac{1-p^{n+1} q^{m+1}/z}{1-p^n q^m z},
\ee
Where $\{z_i\}$ with $i=1,...,N_c$ are the fugacities parameterizing the Cartan subalgebra of $SU(N_c)$, with $\prod_{i=1}^{N_c} z_i = 1$. In addition, in many occasions we will use the shorten notation
\be
\Gamma_e \left(u z^{\pm n} \right)=\Gamma_e \left(u z^{n} \right)\Gamma_e \left(u z^{-n} \right).
\ee

In a similar manner we can write the full contribution of the vector multiplet in the adjoint of $SU(N_c)$, together with the matching Haar measure and projection to gauge singlets as
\be
\frac{\kappa^{N_c-1}}{N_c !} \oint_{\mathbb{T}^{N_c-1}} \prod_{i=1}^{N_c-1} \frac{dz_i}{2\pi i z_i} \prod_{k\ne \ell} \frac1{\Gamma_e(z_k/z_\ell)}\cdots,
\ee
where the dots denote that it will be used in addition to the full matter multiplets transforming in representations of the gauge group. The integration is a contour integration over the maximal torus of the gauge group. $\kappa$ is is the index of $U(1)$ free vector multiplet defined as
\be
\kappa \triangleq (p;p)(q;q),
\ee
where
\be
(a;b) \triangleq \prod_{n=0}^\infty \left( 1-ab^n \right)
\ee
is the q-Pochhammer symbol.

\section{Class $\mathcal{S}_k$ flows of theories with punctures}
\label{A:Flows punc}
Before discussing the possible complications expected with punctures we refer the reader to Appendix \ref{A:PuncConventions} to read on the puncture conventions for class $\mathcal{S}_k$. In the case of theories described by a Riemann surface with punctures some complications are expected. For once, the number of colors for maximal punctures will be lower; thus, one can expect some of the maximal punctures will flow to maximal punctures of some color depending on the chosen vevs, while other will flow to some unknown object and not to a maximal puncture. This will require the initial UV theory to hold only some colors of maximal punctures and not others.

In the case of intermediate punctures (between maximal and minimal) we can expect some mapping between different intermediate punctures. For example in the extreme case of flows from class $\mathcal{S}_2$ to class $\mathcal{S}$ with $N=2$ maximal punctures flow to maximal punctures of class $\mathcal{S}$ that are equivalent to minimal punctures. In addition, in the general case one should expect the color mapping problem mentioned above for these intermediate punctures as well.

The fluxes of the IR theory should get similar expressions only with some additional contribution from punctures. The genus is expected to remain the same as we found without punctures, and the number of punctures should remain the same for the vevs in equation \eqref{E:General vevs} that do not generate additional punctures. For the minimal vevs appearing in \eqref{E:General vevs min} that do generate extra minimal punctures we can expect that their number will additionally depend on the number and type of punctures in the UV theory.

\subsection*{Explicit index example}
In this part we show an explicit example of a punctured theory flow. This will be presented as before via index computations. The vevs we will use to initiate the flow will be the baryon vev as before that for $N=k=2$ is given by
\be
\label{E:vev N=k=2 2}
\left(\beta^{-1}\gamma\left(\frac{pq}{t}\right)\right)^{2}=1\, .
\ee
This translates to the assignments $\beta=\left(\frac{pq}{t}\right)^{1/2}\epsilon^{-1/2}$ and $\gamma=\left(\frac{pq}{t}\right)^{-1/2}\epsilon^{-1/2}$, leaving us with additional minimal punctures of fugacity $\epsilon$. The second baryonic operator vev we will use is
\be
\left(\beta^{-1}\gamma^{-1}\left(\frac{pq}{t}\right)\right)^{2}=1\, ,
\ee
and it will close all additional negative minimal puncture. After the initial flow it translates to
\be
\left(\epsilon\sqrt{\frac{pq}{t}}\right)^{2}=1\, .
\ee

The example we will examine is of the $T_A$ interacting trinion with flipped $\gamma$ flux and flipped maximal puncture colors flow. The index for $T_A$ appeared in the appendix of \cite{Razamat:2016dpl}, and flipping its $\gamma$ flux  and maximal puncture colors is given by simply exchanging $\gamma\to -\gamma$
\be
\mathcal{I}_{T_{A}}(\gamma\to -\gamma) & = & \mathcal{I}_{g=0,m^-=0,s_{1}=3,s_{2}=0,\left(e=1/2,b=-1/4,c=1/4\right)}^{N=2,k=2}\nonumber\\
 & = & \Gamma_{e}\left(t\left(\frac{v_2}{\beta\gamma}\right)^{\pm1}v_{1}^{\pm1}\right)\Gamma_{e}\left(pq\frac{\gamma^2}{\beta^{2}}\right)\kappa\oint\frac{dz}{4\pi iz}\frac{\Gamma_{e}\left(\frac{pq\gamma}{t\beta}\left(\beta\gamma v_{2}^{-1}\right)^{\pm1}z^{\pm1}\right)}{\Gamma_{e}\left(z^{\pm2}\right)}\times\nonumber\\
 & & \Gamma_{e}\left(\gamma^{-1}\beta z^{\pm1}v_{1}^{\pm1}\right)\mathcal{I}_{\mathbf{c},\mathbf{w},\sqrt{zv_{2}},\sqrt{v_{2}/z}}^{orbifold}(\gamma\to -\gamma)\, ,
\ee
where $s_i$ are the number of punctures of color $i$. We remind the reader that the orbifold theory is of two free trinions glued by $\Phi$-gluing to a sphere of two maximal punctures of color 2 and two minimal punctures. The index for the orbifold theory is reproduced here for the readers convenience
\be
\mathcal{I}_{\mathbf{z},\mathbf{c},a,b}^{orbifold} & = & \kappa^{2}\oint\frac{dw_{1}}{4\pi iw_{1}}\oint\frac{dw_{2}}{4\pi iw_{2}}\frac{\Gamma_{e}\left(\frac{pq}{t}\left(\beta\gamma\right)^{\pm1}w_{1}^{\pm1}w_{2}^{\pm1}\right)}{\Gamma_{e}\left(w_{1}^{\pm2}\right)\Gamma_{e}\left(w_{2}^{\pm2}\right)}\times\nonumber\\
 & & \Gamma_{e}\left(t^{\frac{1}{2}}\beta a^{-1}w_{1}^{\pm1}z_{1}^{\pm1}\right)\Gamma_{e}\left(t^{\frac{1}{2}}\gamma^{-1}aw_{1}^{\pm1}z_{2}^{\pm1}\right)\Gamma_{e}\left(t^{\frac{1}{2}}\gamma aw_{2}^{\pm1}z_{1}^{\pm1}\right)\times\nonumber\\
 & & \Gamma_{e}\left(t^{\frac{1}{2}}\beta^{-1}a^{-1}w_{2}^{\pm1}z_{2}^{\pm1}\right)\Gamma_{e}\left(t^{\frac{1}{2}}\gamma bw_{1}^{\pm1}c_{1}^{\pm1}\right)\Gamma_{e}\left(t^{\frac{1}{2}}\beta^{-1}b^{-1}w_{1}^{\pm1}c_{2}^{\pm1}\right)\times\nonumber\\
  & & \Gamma_{e}\left(t^{\frac{1}{2}}\beta b^{-1}w_{2}^{\pm1}c_{1}^{\pm1}\right)\Gamma_{e}\left(t^{\frac{1}{2}}\gamma^{-1}bw_{2}^{\pm1}c_{2}^{\pm1}\right)\, .
\ee

This theory has the required pole to generate the initial flow, but lacks the second vev that effectively closes the negative minimal punctures. In this flow two of the three $SU(2)$ gauge symmetries are Higgsed making several fields massive and decouple in the flow. The index of the IR theory is
\be
\mathcal{I}_{T_{A}}^{flow\ 1}\left(\gamma\to \gamma^{-1}\right) & = & \kappa\oint\frac{dw_{1}}{4\pi iw_{1}}\frac{\Gamma_{e}\left(\frac{pq}{t}\right)^{2}\Gamma_{e}\left(\frac{pq}{t}w_{1}^{\pm2}\right)}{\Gamma_{e}\left(w_{1}^{\pm2}\right)}\nonumber\times\\
 & & \Gamma_{e}\left(\frac{pq}{t}\right)^{-1}\Gamma_{e}\left(t^{1/2}v_{2}^{\pm1}w_{1}^{\pm1}c_{2}^{\pm1}\right)\Gamma_{e}\left(t^{1/2}\epsilon^{\pm1}w_{1}^{\pm1}v_{2}^{\pm1}\right)\nonumber\\
 & = & \Gamma_{e}\left(\frac{pq}{t}\epsilon^{\pm2}\right)^{-1}\mathcal{I}_{g=0,m^{-}=1,s=3,\left(e=1\right)}^{N=2,k=1}=\mathcal{I}_{g=0,s=4,\left(e=1\right)}^{N=2,class\,\mathcal{S}}\, ,
\ee
where $s$ is the number of maximal punctures. The final result appears in two ways for clarity of notation. The first is in the conventions of class $\mathcal{S}_1$ where for $N=2$ minimal and maximal punctures differ by free chirals only. The second is the usual class $\mathcal{S}$ conventions where for $N=2$ we only have one type of punctures. The theory with index $\mathcal{I}_{g=0,s=4,\left(e=1\right)}^{N=2,class\,\mathcal{S}}$ can be interpreted as a $\Phi$-gluing of two free trinions.

ext, we can set the second vev any way if we assume that this theory is glued to another with the required pole. In this case this translates to giving mass to some fields and we find the resulting theory is given by
\be
\mathcal{I}_{T_{A}}^{flow}\left(\gamma\to\gamma^{-1}\right) & = & \kappa\oint\frac{dw_{1}}{4\pi iw_{1}}\frac{\Gamma_{e}\left(\frac{pq}{t}\right)^{2}\Gamma_{e}\left(\frac{pq}{t}w_{1}^{\pm2}\right)}{\Gamma_{e}\left(w_{1}^{\pm2}\right)}\times\nonumber\\
 & & \Gamma_{e}\left(t\right)\Gamma_{e}\left(t^{1/2}v_{2}^{\pm1}w_{1}^{\pm1}c_{2}^{\pm1}\right)\Gamma_{e}\left(\frac{t}{\sqrt{pq}}w_{1}^{\pm1}v_{2}^{\pm1}\right)\nonumber\\
 & = & \mathcal{I}_{g=0,m^{-}=0,s=3,\left(e=3/2\right)}^{N=2,k=1}\, .
\ee
The theory with index $\mathcal{I}_{g=0,m^{-}=0,s=3,\left(e=3/2\right)}^{N=2,k=1}$ can be interpreted as a $\Phi$-gluing of the interacting trinion with a t-flux tube.

We find that in flows that start from theories with punctures, the theory in the end of the flow will depend on the number and properties of the punctures.

\section{Summary of conventions}
\label{A:Conventions}

In this appendix we summarize our various conventions regarding symmetries, fugacities and fluxes used throughout this article. The $6d$ SCFTs that appear in this article generally have an $SU(k)_{\beta} \times SU(k)_{\gamma}\times U(1)_t$ global symmetry, as well as the superconformal symmetry containing the $SU(2)_R$ R-symmetry. The Cartan of these symmetries are generally inherited from the $6d$ theory by the $4d$ theory which is the compactification product. The Cartans of $SU(k)_{\beta} \times SU(k)_{\gamma}$ in $4d$ are denoted as $U(1)_{\beta_i}\times U(1)_{\gamma_i}$ for $i=1,2...,k$, where two are not independent, instead obeying $\sum U(1)_{\beta_i} = \sum U(1)_{\gamma_i} = 0$. In fugacities, these are denoted as $\beta_i$ and $\gamma_i$, obeying $\prod \beta_i = \prod \gamma_i = 1$. In terms of the $SU(k)_{\beta} \times SU(k)_{\gamma}$ $6d$ symmetries, these are defined such that:

\be
\bold{k}_{SU(k)_{\beta}} = \sum^k_{i=1} \beta^N_i, \, \bold{k}_{SU(k)_{\gamma}} = \sum^k_{i=1} \gamma^N_i .
\ee 

The $U(1)_t$ symmetry in $4d$ differs from the 6d $U(1)_t$ by a minus sign, such that $U(1)_t^{6d}=-U(1)_t^{4d}$ and we shall use the fugacity $t$ for it. The flux we consider for $U(1)_t$ is for the 6d symmetry always with no change of sign. This is chosen in order to follow the conventions in the literature.  
Finally we have the Cartan of the $6d$ $SU(2)_R$. This gives a useful, though generically not the superconformal, R-symmetry in $4d$. We shall generically denote it as $U(1)^{6d}_R$, and use the fugacity $r$ for it. We in general don't use it as the superconformal symmetry in index calculations, where the combination $U(1)^{6d}_R - U(1)_t$ is used instead. We generally denote this combination simply as $U(1)_R$, though it should be noted that generically it too is not the superconformal R-symmetry. Nevertheless, whenever a $p q$ appear in a $4d$ index it is with respect to this R-symmetry. We shall also often employ the combination $T=\frac{p q}{t}$.

When compactifying the theory on a Riemann surface we are free to turn on fluxes in the $6d$ flavor symmetries. These fluxes will be denoted by the symbols $(b^{(k)}_i, c^{(k)}_i, e^{(k)})$, where $k$ is that of the $6d$ SCFT. These are defined such that:

\be
\int \frac{F_{\beta_i}}{2\pi} = N b^{(k)}_i, \, \int \frac{F_{\gamma_i}}{2\pi} = N c^{(k)}_i, \, \int \frac{F_{t}}{2\pi} = e^{(k)}.
\ee

Finally, the flux in $U(1)^{6d}_R$ must be proportional to the curvature of the Riemann surface or supersymmetry would not be preserved. This forces:

\be
\int \frac{F_{R_{6d}}}{2\pi} = g-1.
\ee

In this article we shall on several occasions write anomaly polynomials for $4d$ and $6d$ theories, which are written in terms of characteristic classes. Our conventions for the characteristic classes then are as follows. We shall use $p_1 (T)$ and $p_2 (T)$ for the first and second Pontryagin classes of the tangent bundle respectively. For the Chern classes we use the notation $c_n (x)_{\bold{r}}$ the n-th Chern class in the representation of dimension $r$ of the symmetry associated to x. For instance $c_2 (\beta)_{\bold{k}}$ is the second Chern class in the fundamental representation of $SU(k)_{\beta}$. The only exceptions are $c_2 (R)$ and the first Chern classes. For $c_2 (R)$ we shall continue to use the notation introduced in \cite{Ohmori:2014kda}, where it stands for the second Chern class in the doublet representation of $SU(2)_{R}$. First Chern classes are always evaluated with respect to the charge one bundle. For these we use $c_1 (R)$ for the $U(1)^{4d}_R$ first Chern class and $c_1 (R')$ for the $U(1)^{6d}_R$ one. 

\end{appendix}


\bibliographystyle{ytphys}
\bibliography{refs}

\end{document}